\documentclass[twocolumn,aps,floats,amsmath,longbibliography]{revtex4-1}
\usepackage{graphicx}
\usepackage{subfigure}
\usepackage{hyperref} 
\usepackage{bm}
\usepackage{float}
\usepackage{yfonts}
\usepackage{xcolor}
\hypersetup{
    colorlinks,
    linkcolor={blue!80!black},
    citecolor={blue!80!black},
    urlcolor={blue!80!black}
}
\usepackage[mathcal]{eucal}
\usepackage{etoolbox}

\begin{document}

\title{Time scales for charge-transfer based operations on Majorana systems}
\author{R. Seoane Souto$^{1,2}$, K. Flensberg$^{1}$ and M. Leijnse$^{1,2}$}
\date{\today}
\affiliation{$^1$Center for Quantum Devices and Station Q Copenhagen, Niels Bohr Institute, University of Copenhagen, DK-2100 Copenhagen, Denmark\\
$^2$Division of Solid State Physics and NanoLund, Lund University, Box 118, S-22100 Lund, Sweden}
\begin{abstract}
In this article we analyze the efficiency of operations based on transferring charge from a quantum dot (QD) to two coupled topological superconductors, which can be used for performing nonabelian operations on Majorana bound states (MBSs). We develop a method which allows us to describe the full time-evolution of the system as the QD energy is manipulated. Using a full counting statistics analysis, we set bounds to the operation time scales. The lower bound depends on the superconducting phase difference due to a partial decoupling of the different MBSs parity sectors, while the upper bound is set by the tunneling of quasiparticles to the MBSs. Using realistic parameters, we find the existence of a regime where the operation can be carried out with a fidelity close to unity. Finally, we propose the use of a two operations protocol to quantify the effect of the dephasing and accumulated dynamical phases, demonstrating their absence for certain superconducting phase differences.

\end{abstract}
\maketitle
\section{Introduction}

Prediction of the existence of Majorana bound states (MBSs) at the ends of a topological superconductor (TS) has given hope of observing nonabelian statistics (for recent reviews see \cite{Alicea_RPP2012,Leijnse_Review2012,Beenakker_review2013,Aguado_Nuovo2017,Lutchyn_NatRev2018}), which can potentially be used for topological quantum computation \cite{Nayak_RMP2008}. Earliest evidence of their existence were based on local probes coupled to the ends of a TS, showing the build-up of a zero bias peak, firstly measured in Ref. \cite{Mourik_science2012}. More recently, some experiments have shown additional pieces of evidence consistent with the presence of MBSs in proximity-induced superconducting devices such as the zero bias conductance quantization \cite{Nichele_PRL2017,Zhang_Nature2018}, interferometry signatures \cite{Whiticar_2019}, exponential scaling with length in Coulomb blockaded islands \cite{Albrecht_nature2016} and interactions between zero-energy states and quantum dot (QD) orbitals \cite{Deng_Science2016}.
 
However, an unambiguous proof of their topological origin would rely on the demonstration of their nonabelian statistics. Some previous theoretical proposals in this direction were based on a spatial exchange of MBSs in a multiterminal device \cite{Alicea_NatPhys2011} and others used tunneling or Coulomb blockade to manipulate the ground state manifolds \cite{Sau_PRB2011,Hyart_PRB2013,Aasen_PRX2016}. Alternatively, signatures of the predicted nonabelian statistics can emerge after a sequence of manipulations of the quantum state of a set of MBSs, using charge-transfer based operations of QDs coupled to TSs \cite{Flensberg_PRL2011}. The dependence of the final state on the order in which the operations are performed would reveal the nonabelian nature of MBSs.
 
\begin{figure}
	\includegraphics[width=0.45\textwidth]{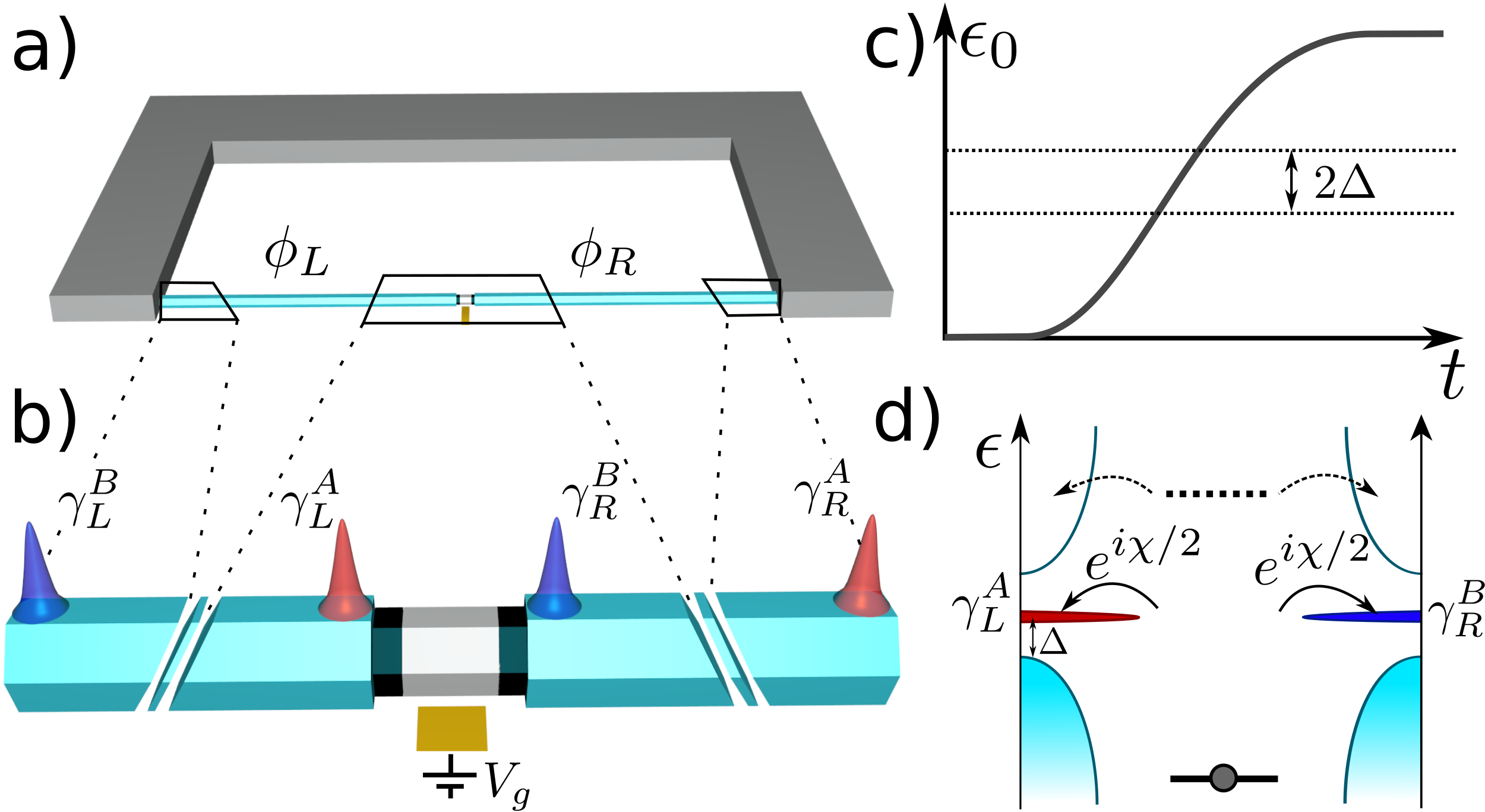}
	\caption{a) Schematic representation of the device consisting of two long TS wires (blue) embedded into a superconducting loop. b) Enlarged representation of the ends of the wires, which host MBSs ($\gamma_{L,R}^{A,B}$), and the QD used for the operations. c) Time evolution of the QD energy level. d) Energy representation, where curved arrows represent the processes of the electron tunneling to the MBSs (solid lines) or to the continuum of states (dashed lines) during the manipulation of the QD level energy (thick horizontal line).}
	\label{Fig1}
\end{figure}

In this work, we analyze the efficiency of the charge-transfer operation of a QD coupled to two TS electrodes \cite{Flensberg_PRL2011}, schematically represented in Figs. \ref{Fig1} a) and b). During the operation, the QD level energy is swept from negative to positive energies well outside the superconducting gap, as illustrated in Fig.  \ref{Fig1} c). As a consequence, the electron initially residing in the QD is transferred to the TSs. In a successful operation the QD empties into the MBSs  (solid arrows in Fig. \ref{Fig1} d)). As an example, starting from an initial $\left|p_L, p_R\right\rangle_i=\left|1,-1\right\rangle$ state (where $\pm1$ means even/odd MBS parity of one of the leads), it evolves to a superposition between $\left|1,1\right\rangle$ and $\left|-1,-1\right\rangle$ after a successful operation, with weights proportional to the tunneling amplitudes to the right and left TSs, respectively,  and a relative phase given by the superconducting phase difference. Differently from Ref. \cite{Flensberg_PRL2011}, we consider the full time dynamics of the system, taking also into account the degrees of freedom from the continuum of states. This allows us to analyze various sources of errors such as the non-adiabatic effects, which can make the QD charge relax (dashed arrows in Fig. \ref{Fig1} d)), and the role of excited quasiparticles above the superconducting gap, which can produce undesired parity changes.

In order to analyze the change in the MBSs parity, we use a full counting statistics formalism which allows us to keep track of the transferred electrons in the junction. For experimentally accessible temperatures, we find an optimal manipulation rate window where operations can be performed with high fidelity. We propose a protocol based on $2$ charge transfer operations, where the QD level returns back to its original value at the final time, as a way of quantifying the influence of dephasing and accumulated dynamical phases. Deviations between the initial and final states provides a measurement of the effect of the different sources of errors.

\section{Theoretical framework}
We consider two TSs coupled to a QD schematically shown in Figs. \ref{Fig1} a) and b) and described by the Hamiltonian
\begin{equation}
	H=H_{L}+H_{R}+H_{QD}+H_{T}\,.
	\label{H_0}
\end{equation}
 Here, $H_{\nu}=\sum_n \left(\hat{\Psi}_{n,\nu}^\dagger \hat{h}_{0,\nu}\hat{\Psi}_{n,\nu}+\hat{\Psi}_{n+1,\nu}^\dagger \hat{h}_{s,\nu}\hat{\Psi}_{n,\nu}\right)$ is the Hamiltonian of the TSs, described by the spinless Kitaev Hamiltonian \cite{Kitaev_2001}, where $\hat{\Psi}_{n,\nu}=\left(c_{n,\nu},c_{n,\nu}^\dagger\right)^T$ is the Nambu spinor, with $c_{n,\nu}$ ($c_{n,\nu}^\dagger$) being the annihilation (creation) operators on site $n$ of electrode $\nu=L,R$. The first term describes the on-site energy, $\hat{h}_{0,\nu}=-\mu\hat{\tau}_z/2$. The second term contains information about the hopping between neighboring sites and p-wave superconducting correlations, $\hat{h}_{s,\nu}=-t_0\hat{\tau}_z/2+i\Delta\hat{\tau}_y/2$, with $t_0$ the normal hopping between neighbors and $\Delta$ the induced superconducting gap, taken to be the same in both electrodes for simplicity. The Hamiltonian of the spinless QD is given by $H_{QD}=\hat{\Psi}^{\dagger}_{d} \hat{h}_0\hat{\Psi}_d$, where $\hat{h}_0=\epsilon_0(t)\hat{\tau}_z/2$ and $\hat{\Psi}_{d}=\left(d,d^\dagger\right)^T$. The time-dependent QD level energy, $\epsilon_0(t)$, evolves during the considered operation from the initial value $\epsilon_i$ to the final one $\epsilon_f$ (with $|\epsilon_i|,|\epsilon_f|\gg\Delta$). For simplicity, $\epsilon_0(t)$ is assumed to evolve linearly in time close to the superconducting gap with a slope  $r$, as illustrated in Fig. \ref{Fig1} c). Finally, $H_{T}=\sum_{\nu}(\hat{\Psi}^{\dagger}_{0,\nu}\hat{h}_{T_\nu}\hat{\Psi}_{d}+\mbox{H.C.})$ describes the tunneling of electrons between the QD and the closest site of the two TSs with $\hat{h}_{T_\nu}=t_\nu\hat{\tau}_z e^{i\hat{\tau}_z\phi_\nu/2}$, where $\phi=\phi_L-\phi_R$ determines the superconducting phase difference between the leads. The tunneling rates are defined as $\Gamma_\nu=\pi(t_\nu)^2\rho_F$, with $\Gamma=\Gamma_L+\Gamma_R$ and $\rho_F$ being the normal density of states at the Fermi level, taken as $\rho_F=1/\Delta$.
 
 The dynamics of the system during an adiabatic operation is accurately described by a low energy model, which only considers the MBSs in the leads. It is described by the Hamiltonian
 \begin{equation}
  H_{LEM}=H_{QD}+H_{T,MBS}\,,
  \label{H_LEM}
 \end{equation}
where $ H_{T,MBS}=V_L\gamma_{L}^Ae^{i\phi_L/2}d+V_R\gamma_{R}^Be^{i\phi_R/2}d+\mbox{H.C.}$ is the tunneling Hamiltonian, with $V_\nu=t_\nu\rho_F\Delta$. Here, $\gamma_{L}^{A}=c_{0,L}+c_{0,L}^\dagger$ and $\gamma_{R}^{B}=i(c_{0,R}-c_{0,R}^\dagger)$ are the MBS operators at the closest end to the QD of the left and right TS (see Fig. \ref{Fig1} b)).

We use the non-equilibrium Green function (NEGF) formalism to access the system state. In the time-dependent regime, the inverse QD NEGF is given by 
\begin{equation}
	G^{-1}(t,t')=g_{0}^{-1}(t,t')-\Sigma(t,t')\,,
	\label{g_0-1}
\end{equation}
where $g^{-1}_0$ represents the inverse QD Green function in the Keldysh-Nambu space \cite{Kamenev_2011,Souto_PRL2016,Souto_PRB2017} and $\Sigma$ the tunneling self-energy. The self-energy can be decomposed into two contributions, $\Sigma=\Sigma_M+\Sigma_c$, describing the coupling to the MBSs and the continuum of states, respectively (expressions are given in the supplementary information (SI) \cite{SM}). For simplicity, we consider a sudden connection between the QD and the electrodes at $t=0$, such that the charge state of the QD and the parity state of the electrodes are initially well defined. We have checked that the connection process does not affect to the initial parity state of the MBSs for an initial level position $|\epsilon_i|\gg\Delta$.

The problem is solved in the time domain by discretizing the Keldysh contour and numerically inverting Eq. (\ref{g_0-1}). Information about the mean properties of the system can be extracted from the QD NEGFs. For instance, the average parity of each MBS can be obtained as $\left\langle p_\nu\right\rangle=i\left\langle \gamma_{\nu}^{A}\gamma_{\nu}^{B}\right\rangle$  (details are provided in the SI \cite{SM}).

In contrast, from the mean values obtained from the NEGFs of Eq. (\ref{g_0-1}) it is not possible to extract the probability of a successful operation. This information can instead be obtained by projecting the number of electrons transferred between the QD and the MBSs using a full counting statistics analysis. This information is encoded in the generating function
\begin{equation}
	Z(\chi,t)=\left\langle T_C \exp\left[-i\int_C dt'\sum_\nu H_{T,\chi}(t')\right]\right\rangle_0\,,
\end{equation}
where the average is taken over the decoupled system and the counting field, $\chi$, is a phase factor used to count tunneling electrons. It enters into the tunneling amplitudes as $\hat{h}_{T_\nu,\chi_\nu}=\hat{h}_{T_\nu}e^{i\tau_{z}^K(\hat{\tau}_0+\hat{\tau}_z)\chi_\nu/2}$, where $\tau_{z}^K$ is the Pauli matrix in Keldysh space, indicating the change on the counting field sign between the two Keldysh branches. We consider the counting field acting only for electrons tunneling between the QD and the MBSs, as illustrated in Fig \ref{Fig1} d) (and not between the QD and the quasiparticle states outside the gap), and take $\chi_L=\chi_R=\chi/2$, which projects the wavefunction onto a state with well defined joint parity of left and right MBSs.

The GF can also be written as $Z(\chi,t)=\sum_n \mathcal{P}_n(t)e^{in\chi}$, where $\mathcal{P}_n$ is the probability of transferring $n$ charges between the QD and the MBSs after a time $t$. Note that at $\chi=\pi$ the GF describes the MBSs parity change. In the  Keldysh-Nambu space, the GF can be computed as a Fredholm determinant \cite{Kamenev_2011,Souto_PRL2016}
\begin{equation}
	Z(\chi,t)=\frac{\mbox{det}[g^{-1}_0-\Sigma(\chi)]}{\mbox{det}[g^{-1}_0-\Sigma(\chi=0)]}\,.
\end{equation}
The probability of a change of the joint MBS parity is described by the generating function as $P(t)=[1-Z(\chi=\pi,t)]/2$, which is also the probability for the QD charge to be transferred to the MBSs. $P$ saturates for $|\epsilon_0(t)|\sim10\Gamma$ to a value $P_f$, leading to a total operation time $t\approx20\,\Gamma/r$.

\section{Results}
\label{sec::results}
\begin{figure}
	\includegraphics[width=0.5\textwidth]{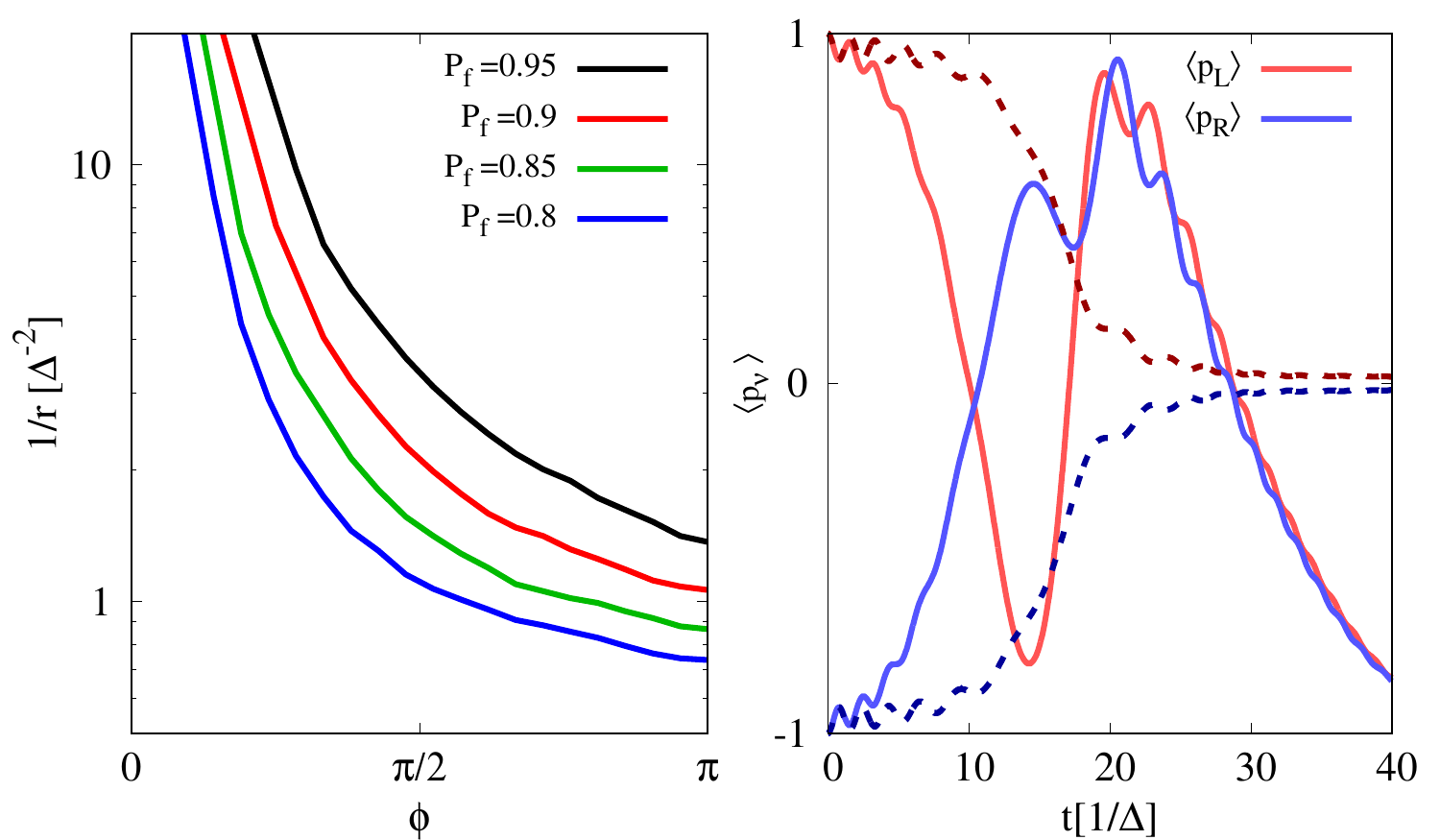}
	\caption{Left panel: Dependence of the inverse of the operation rate as a function of the superconducting phase for different $P_f$ values. Right panel: average parity of left and right MBSs during an operation with $r=\Delta^2/4$. We consider two different phases: $\phi=\pi/2$ (solid lines) and $\phi=\pi$ (dotted lines). In both cases we take $\Gamma_L=\Gamma_R=0.15\Delta$ and temperature $T=0$.}
	\label{Fig2}
\end{figure}
In this section we analyze the success probability of the charge-transfer based operation on the system sketched in Figs. \ref{Fig1} a) and b). We consider an initially well-defined MBS parity on both sides, choosing without loss of generality $\left| p_L, p_R\right\rangle_i=\left| 1,-1\right\rangle$ as the initial MBS parity state. A successful operation in which the QD charge is efficiently transferred to the MBSs (solid arrows in Fig. \ref{Fig1} d)) leads to a change of the total MBS parity state (from odd to even in our case), leading to a final state $\left| p_L, p_R\right\rangle_f=(V_Re^{i\phi_R/2}\left| 1,1\right\rangle+V_Le^{i\phi_L/2}\left| -1,-1\right\rangle)/\sqrt{V_{L}^2+V_{R}^2}$. As discussed below, this is only true for $\phi=(2n+1)\pi$.

In the left panel of Fig. \ref{Fig2} we show the dependence of the inverse manipulation rate ($1/r$) on the superconducting phase difference for different values of $P_f$. As shown, the rate needed for a given $P_f$ depends strongly on $\phi$, becoming optimal (thus, minimizing operation times) for $\phi=(2n+1)\pi$. At $\phi=2n\pi$ the time of the operation diverges due to the formation of a dark state preserving the total MBS parity state for $V_L=V_R$ and for any $\epsilon_0$ value, setting a limit $P_f=0.5$ for adiabatic operations. The dark state is given by $(\left|1,-1 \right\rangle+i\left|-1,1 \right\rangle)/\sqrt{2}$ for the odd MBS parity sector and $(\left|1,1 \right\rangle+i\left|-1,-1 \right\rangle)/\sqrt{2}$ for the even one. For $V_L\neq V_R$ the even and odd MBS parity sectors are no longer decoupled, although the value of $r$ needed for reaching a $P_f$ close to unity becomes strongly dependent on the asymmetry between the left and right tunneling rates close to $\phi=2n\pi$. 

In the right panel of Fig. \ref{Fig2} we show the average parity of the left and right MBSs for an adiabatic $r$ value where $P_f\approx1$ and two different superconducting phases. As shown, for $\phi=\pi$ (mod $2\pi$) $\left\langle p_L\right\rangle$ and $\left\langle p_R\right\rangle$ tend to $0$ at long times, consistent with the convergence to the expected final state after the operation. In contrast, for $\phi\neq(2n+1)\pi$ the average parities exhibit oscillations during the manipulation, which leads to a dependence of the final state on the details of the operation. These oscillations are  due to the breaking of degeneracy between the even and odd parity states of the non-local fermion formed by the left and the right MBSs, as shown in the SI \cite{SM}. For $\phi=(2n+1)\pi+\delta\phi$ the period of the oscillations is given by $\sqrt{\epsilon^{2}_0+4(V^{2}_L+V^{2}_R)}/(4V_L V_R \delta\phi)$ with $\delta \phi\ll1$. This illustrates the importance of having an accurate control on the superconducting phase difference. Additional oscillations are observed after the operation (for $|\epsilon_0|\gg\Delta$) with a period given by $\epsilon_0/[4V_LV_R\cos(\phi/2)]$. It is worth commenting that the same behavior is observed in the case the QD is coupled to the same Majorana operators (either $\gamma^A$ or $\gamma^B$) at the left and right side, as considered in Ref. \cite{Flensberg_PRL2011}, by shifting the superconducting phase difference value by $\pi$. In the following we will focus on the $\phi=(2n+1)\pi$ situation, where the parity oscillations are absent and $P_f$ quantifies the success probability of the charge-transfer operation.

\begin{figure}
	\includegraphics[width=0.5\textwidth]{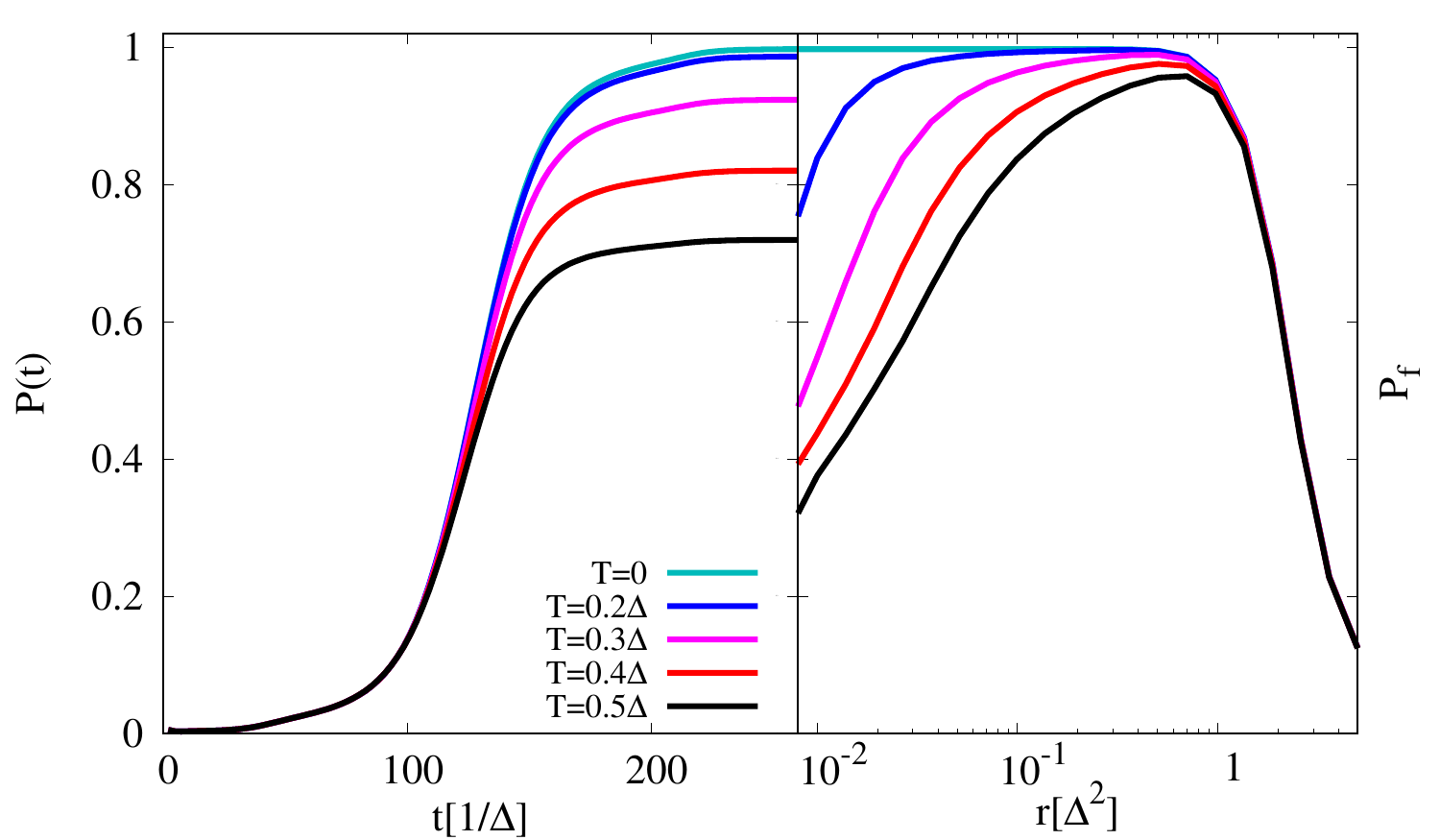}
	\caption{Temperature dependence of the MBSs parity change probability, showing the time evolution (left panel) and the long time limit as a function of the rate (right panel). Parameters are the same as in Fig. \ref{Fig2} with $\phi=\pi$.}
	\label{Fig3}
\end{figure}

The analysis performed above would indicate that the choice of slower rates tends to improve the success probability. However, some additional processes can affect the operation fidelity for adiabatic rates, such as the decaying of excited quasiparticles into the MBSs, which lead to uncontrolled parity changes. First estimations have shown that these events happen in the ms time scale in a trivial wire, i.e. without MBSs \cite{Higginbotham_NatPhys2015}. Processes involving the decaying of quasiparticles directly into the MBSs will be disregarded in the following. Instead we consider the time-scales for quasiparticles tunneling between the continuum and the MBSs mediated by the QD during the charge-transfer based operation which, as it turns out,  can be much shorter, see Fig. \ref{Fig3}. At a finite $T$, there is a density of quasiparticles excited above the superconducting gap. For small $T$, the number of quasiparticles is exponentially suppressed and the manipulation can be performed with a high success probability, as illustrated by the blue curves in the left panel of Fig. \ref{Fig3}. However, when the temperature increases, $P_f$ drops for adiabatic rates. This is better illustrated in the right panel of Fig. \ref{Fig3} where we show that $P_f$ exhibits a maximum, decreasing for small $r$ values, due to the tunneling of thermally excited quasiparticles to the QD.

For small temperatures ($T\lesssim0.2\Delta$), the system exhibits a range of $r$ values where $P_f$ is close to unity. We observe that the decay point of $P_f$ for decreasing $r$ values scales linearly with the density of thermally excited quasiparticles. It allows us to estimate an optimal window for the rate using experimentally relevant parameters where $k_BT/\Delta\sim0.1$, finding $r/\Delta\Gamma\sim10^{-4}-10^{-1}$. This leads to an estimated manipulation time scale of the order of $0.1\,\mu\mbox{s}$ to $0.1\,\mbox{ns}$ for $V_L$ and $V_R$ smaller but of the same order of magnitude as $\Delta$. The upper bound has been estimated considering an equilibrium distribution of excited quasiparticles with energies larger than the superconducting gap. It can be significantly reduced if a higher density of these quasiparticles is found in the TSs.

\begin{figure}
	\includegraphics[width=0.5\textwidth]{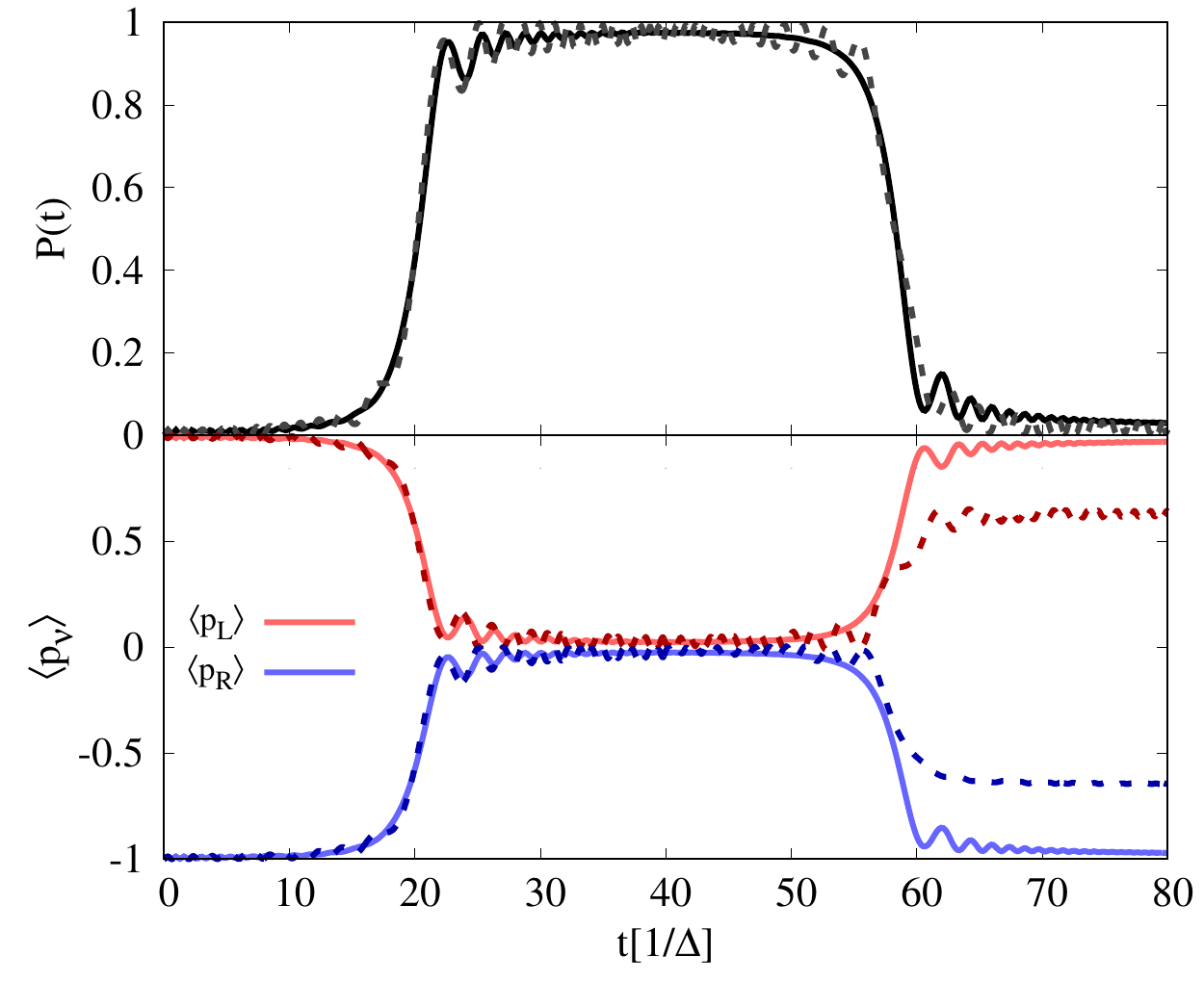}
	\caption{Results for a double charge-transfer operations where the QD level returns to its initial value at the final time. The solid curves describe the coherent dynamics while the dashed ones illustrate the effect of a randomly distributed phase, $\xi \in[-\pi/2,\pi/2]$, introduced in the quantum state after the first manipulation, using the low energy Hamiltonian of Eq. (\ref{H_LEM}). Upper panel: probability of total MBS parity change. Lower panel: left and right MBS parity. The parameters are the same as in Fig. \ref{Fig2} with $\phi=\pi$ and $r=\Delta^2/2$.}
	\label{Fig4}
\end{figure}
 
 While Figs. \ref{Fig2} and \ref{Fig3} describe the probability of changing the parity of the MBSs, they do not contain information about dephasing or the accumulation of dynamical phases. These effects can be generated by high order processes involving the tunneling of electrons between the QD and the continuum of states. In order to quantify the dephasing, we suggest two charge-transfer operations, where the QD level energy is swept from $\epsilon_i$ to $\epsilon_f$ and then back to $\epsilon_i$. The results of this protocol are shown in Fig. \ref{Fig4} for a situation where a high $P_f$ is observed in the right panel of Fig. \ref{Fig3} (solid lines). In the top panel of Fig. \ref{Fig4} we show the change of the joint MBS parity, which evolves to almost $1$ after the first manipulation and returns back to almost $0$ when the QD level energy returns to its original value (solid line). The lower panel of Fig. \ref{Fig4} shows the average parity values of the left and right MBS pairs during the operation, confirming that the parity of each electrode returns back to its original value after the second operation (solid lines). We have also verified that the final state of the electrodes does not degrade if different adiabatic rates are considered for the two manipulation processes, indicating the absence of uncontrolled dynamical phases for $\phi=\pi$ (mod $2\pi$).
  
 The dashed lines in Fig. \ref{Fig4} illustrate the effect of an accumulation of uncontrolled phase factors after the first operation, which we simulate using a randomly distributed phase $\xi$ in a given interval. Therefore, the state after a successful first operation is $\propto(V_Re^{i\phi_R/2}\left| 1,1\right\rangle+V_Le^{i(\phi_L/2+\xi)}\left| -1,-1\right\rangle)$. This is analyzed using the low energy model described in Eq. (\ref{H_LEM}), which accurately describes the system adiabatic evolution. While the total MBS parity change is not affected by the inclusion of randomly distributed phase $\xi$ (top panel of Fig. \ref{Fig4}), the individual parities of the electrodes do not return back to their original values. Instead, they they approach $\left\langle p_\nu\right\rangle=0$ in the limit the introduced phase factor is completely random. Therefore, the difference between the initial and final parity states, $\delta(\left\langle p_\nu\right\rangle)=\left\langle p_\nu\right\rangle_f-\left\langle p_\nu\right\rangle_i$, quantifies the accumulated error in the whole process. The error on a single manipulation can be estimated as $1-\sqrt{1-|\delta (\left\langle p_\nu\right\rangle)|}$ in the regime where no other errors occur between the first and the second operation.
 
\section{Conclusions}
We have analyzed the efficiency of the charge-transfer based operations of a QD coupled to two topological superconductors, setting bounds to its optimal time scales. We have found the optimal parameters where the final state after an adiabatic operation does not depend on the details of how it is performed. We have also analyzed the effects of imperfections in the system that affect the adiabatic operations, studying the effect of the excited quasiparticles above the superconducting gap at finite temperature, which leads to a lower bound to the manipulation rate. In the SI \cite{SM} we furthermore investigate the effects of an overlap between the MBS wavefunctions within a wire. We show that this leads to an upper bound on the operation time scale determined by the inverse of the difference between the overlap strength on both sides. Our results establish charge-transfer based operations as a realistic alternative to experimentally probe the nonabelian nature of MBSs.

\section{Acknowledgements}
We acknowledge valuable contributions from A. Levy Yeyati and A. Mart\'in-Rodero on the code development. We acknowledge funding from the Danish National Research Foundation. RSS and ML acknowledge funding from QuantERA project “2D hybrid materials as a platform for topological quantum computing” and from NanoLund.

\bibliographystyle{apsrev4-1}

\bibliography{bibliography.bib}{}

\begin{thebibliography}{26}%
\makeatletter
\providecommand \@ifxundefined [1]{%
 \@ifx{#1\undefined}
}%
\providecommand \@ifnum [1]{%
 \ifnum #1\expandafter \@firstoftwo
 \else \expandafter \@secondoftwo
 \fi
}%
\providecommand \@ifx [1]{%
 \ifx #1\expandafter \@firstoftwo
 \else \expandafter \@secondoftwo
 \fi
}%
\providecommand \natexlab [1]{#1}%
\providecommand \enquote  [1]{``#1''}%
\providecommand \bibnamefont  [1]{#1}%
\providecommand \bibfnamefont [1]{#1}%
\providecommand \citenamefont [1]{#1}%
\providecommand \href@noop [0]{\@secondoftwo}%
\providecommand \href [0]{\begingroup \@sanitize@url \@href}%
\providecommand \@href[1]{\@@startlink{#1}\@@href}%
\providecommand \@@href[1]{\endgroup#1\@@endlink}%
\providecommand \@sanitize@url [0]{\catcode `\\12\catcode `\$12\catcode
  `\&12\catcode `\#12\catcode `\^12\catcode `\_12\catcode `\%12\relax}%
\providecommand \@@startlink[1]{}%
\providecommand \@@endlink[0]{}%
\providecommand \url  [0]{\begingroup\@sanitize@url \@url }%
\providecommand \@url [1]{\endgroup\@href {#1}{\urlprefix }}%
\providecommand \urlprefix  [0]{URL }%
\providecommand \Eprint [0]{\href }%
\providecommand \doibase [0]{http://dx.doi.org/}%
\providecommand \selectlanguage [0]{\@gobble}%
\providecommand \bibinfo  [0]{\@secondoftwo}%
\providecommand \bibfield  [0]{\@secondoftwo}%
\providecommand \translation [1]{[#1]}%
\providecommand \BibitemOpen [0]{}%
\providecommand \bibitemStop [0]{}%
\providecommand \bibitemNoStop [0]{.\EOS\space}%
\providecommand \EOS [0]{\spacefactor3000\relax}%
\providecommand \BibitemShut  [1]{\csname bibitem#1\endcsname}%
\let\auto@bib@innerbib\@empty
\bibitem [{\citenamefont {Alicea}(2012)}]{Alicea_RPP2012}%
  \BibitemOpen
  \bibfield  {author} {\bibinfo {author} {\bibfnamefont {J.}~\bibnamefont
  {Alicea}},\ }\href {\doibase 10.1088/0034-4885/75/7/076501} {\bibfield
  {journal} {\bibinfo  {journal} {Reports on Progress in Physics}\ }\textbf
  {\bibinfo {volume} {75}},\ \bibinfo {pages} {076501} (\bibinfo {year}
  {2012})}\BibitemShut {NoStop}%
\bibitem [{\citenamefont {Leijnse}\ and\ \citenamefont
  {Flensberg}(2012)}]{Leijnse_Review2012}%
  \BibitemOpen
  \bibfield  {author} {\bibinfo {author} {\bibfnamefont {M.}~\bibnamefont
  {Leijnse}}\ and\ \bibinfo {author} {\bibfnamefont {K.}~\bibnamefont
  {Flensberg}},\ }\href {\doibase 10.1088/0268-1242/27/12/124003} {\bibfield
  {journal} {\bibinfo  {journal} {Semiconductor Science and Technology}\
  }\textbf {\bibinfo {volume} {27}},\ \bibinfo {pages} {124003} (\bibinfo
  {year} {2012})}\BibitemShut {NoStop}%
\bibitem [{\citenamefont {Beenakker}(2013)}]{Beenakker_review2013}%
  \BibitemOpen
  \bibfield  {author} {\bibinfo {author} {\bibfnamefont {C.}~\bibnamefont
  {Beenakker}},\ }\href {\doibase 10.1146/annurev-conmatphys-030212-184337}
  {\bibfield  {journal} {\bibinfo  {journal} {Annual Review of Condensed Matter
  Physics}\ }\textbf {\bibinfo {volume} {4}},\ \bibinfo {pages} {113} (\bibinfo
  {year} {2013})}\BibitemShut {NoStop}%
\bibitem [{\citenamefont {Aguado}(2017)}]{Aguado_Nuovo2017}%
  \BibitemOpen
  \bibfield  {author} {\bibinfo {author} {\bibfnamefont {R.}~\bibnamefont
  {Aguado}},\ }\href {\doibase 10.1393/ncr/i2017-10141-9} {\bibfield  {journal}
  {\bibinfo  {journal} {Riv. Nuovo Cimento}\ }\textbf {\bibinfo {volume}
  {40}},\ \bibinfo {pages} {523} (\bibinfo {year} {2017})}\BibitemShut
  {NoStop}%
\bibitem [{\citenamefont {Lutchyn}\ \emph {et~al.}(2018)\citenamefont
  {Lutchyn}, \citenamefont {Bakkers}, \citenamefont {Kouwenhoven},
  \citenamefont {Krogstrup}, \citenamefont {Marcus},\ and\ \citenamefont
  {Oreg}}]{Lutchyn_NatRev2018}%
  \BibitemOpen
  \bibfield  {author} {\bibinfo {author} {\bibfnamefont {R.~M.}\ \bibnamefont
  {Lutchyn}}, \bibinfo {author} {\bibfnamefont {E.~P. A.~M.}\ \bibnamefont
  {Bakkers}}, \bibinfo {author} {\bibfnamefont {L.~P.}\ \bibnamefont
  {Kouwenhoven}}, \bibinfo {author} {\bibfnamefont {P.}~\bibnamefont
  {Krogstrup}}, \bibinfo {author} {\bibfnamefont {C.~M.}\ \bibnamefont
  {Marcus}}, \ and\ \bibinfo {author} {\bibfnamefont {Y.}~\bibnamefont
  {Oreg}},\ }\href {\doibase 10.1038/s41578-018-0003-1} {\bibfield  {journal}
  {\bibinfo  {journal} {Nature Reviews Materials}\ }\textbf {\bibinfo {volume}
  {3}},\ \bibinfo {pages} {52} (\bibinfo {year} {2018})}\BibitemShut {NoStop}%
\bibitem [{\citenamefont {Nayak}\ \emph {et~al.}(2008)\citenamefont {Nayak},
  \citenamefont {Simon}, \citenamefont {Stern}, \citenamefont {Freedman},\ and\
  \citenamefont {Das~Sarma}}]{Nayak_RMP2008}%
  \BibitemOpen
  \bibfield  {author} {\bibinfo {author} {\bibfnamefont {C.}~\bibnamefont
  {Nayak}}, \bibinfo {author} {\bibfnamefont {S.~H.}\ \bibnamefont {Simon}},
  \bibinfo {author} {\bibfnamefont {A.}~\bibnamefont {Stern}}, \bibinfo
  {author} {\bibfnamefont {M.}~\bibnamefont {Freedman}}, \ and\ \bibinfo
  {author} {\bibfnamefont {S.}~\bibnamefont {Das~Sarma}},\ }\href {\doibase
  10.1103/RevModPhys.80.1083} {\bibfield  {journal} {\bibinfo  {journal} {Rev.
  Mod. Phys.}\ }\textbf {\bibinfo {volume} {80}},\ \bibinfo {pages} {1083}
  (\bibinfo {year} {2008})}\BibitemShut {NoStop}%
\bibitem [{\citenamefont {Mourik}\ \emph {et~al.}(2012)\citenamefont {Mourik},
  \citenamefont {Zuo}, \citenamefont {Frolov}, \citenamefont {Plissard},
  \citenamefont {Bakkers},\ and\ \citenamefont
  {Kouwenhoven}}]{Mourik_science2012}%
  \BibitemOpen
  \bibfield  {author} {\bibinfo {author} {\bibfnamefont {V.}~\bibnamefont
  {Mourik}}, \bibinfo {author} {\bibfnamefont {K.}~\bibnamefont {Zuo}},
  \bibinfo {author} {\bibfnamefont {S.~M.}\ \bibnamefont {Frolov}}, \bibinfo
  {author} {\bibfnamefont {S.~R.}\ \bibnamefont {Plissard}}, \bibinfo {author}
  {\bibfnamefont {E.~P. A.~M.}\ \bibnamefont {Bakkers}}, \ and\ \bibinfo
  {author} {\bibfnamefont {L.~P.}\ \bibnamefont {Kouwenhoven}},\ }\href
  {\doibase 10.1126/science.1222360} {\bibfield  {journal} {\bibinfo  {journal}
  {Science}\ }\textbf {\bibinfo {volume} {336}},\ \bibinfo {pages} {1003}
  (\bibinfo {year} {2012})}\BibitemShut {NoStop}%
\bibitem [{\citenamefont {Nichele}\ \emph {et~al.}(2017)\citenamefont
  {Nichele}, \citenamefont {Drachmann}, \citenamefont {Whiticar}, \citenamefont
  {O'Farrell}, \citenamefont {Suominen}, \citenamefont {Fornieri},
  \citenamefont {Wang}, \citenamefont {Gardner}, \citenamefont {Thomas},
  \citenamefont {Hatke}, \citenamefont {Krogstrup}, \citenamefont {Manfra},
  \citenamefont {Flensberg},\ and\ \citenamefont {Marcus}}]{Nichele_PRL2017}%
  \BibitemOpen
  \bibfield  {author} {\bibinfo {author} {\bibfnamefont {F.}~\bibnamefont
  {Nichele}}, \bibinfo {author} {\bibfnamefont {A.~C.~C.}\ \bibnamefont
  {Drachmann}}, \bibinfo {author} {\bibfnamefont {A.~M.}\ \bibnamefont
  {Whiticar}}, \bibinfo {author} {\bibfnamefont {E.~C.~T.}\ \bibnamefont
  {O'Farrell}}, \bibinfo {author} {\bibfnamefont {H.~J.}\ \bibnamefont
  {Suominen}}, \bibinfo {author} {\bibfnamefont {A.}~\bibnamefont {Fornieri}},
  \bibinfo {author} {\bibfnamefont {T.}~\bibnamefont {Wang}}, \bibinfo {author}
  {\bibfnamefont {G.~C.}\ \bibnamefont {Gardner}}, \bibinfo {author}
  {\bibfnamefont {C.}~\bibnamefont {Thomas}}, \bibinfo {author} {\bibfnamefont
  {A.~T.}\ \bibnamefont {Hatke}}, \bibinfo {author} {\bibfnamefont
  {P.}~\bibnamefont {Krogstrup}}, \bibinfo {author} {\bibfnamefont {M.~J.}\
  \bibnamefont {Manfra}}, \bibinfo {author} {\bibfnamefont {K.}~\bibnamefont
  {Flensberg}}, \ and\ \bibinfo {author} {\bibfnamefont {C.~M.}\ \bibnamefont
  {Marcus}},\ }\href {\doibase 10.1103/PhysRevLett.119.136803} {\bibfield
  {journal} {\bibinfo  {journal} {Phys. Rev. Lett.}\ }\textbf {\bibinfo
  {volume} {119}},\ \bibinfo {pages} {136803} (\bibinfo {year}
  {2017})}\BibitemShut {NoStop}%
\bibitem [{\citenamefont {Zhang}\ \emph {et~al.}(2018)\citenamefont {Zhang},
  \citenamefont {Liu}, \citenamefont {Gazibegovic}, \citenamefont {Xu},
  \citenamefont {Logan}, \citenamefont {Wang}, \citenamefont {van Loo},
  \citenamefont {Bommer}, \citenamefont {de~Moor}, \citenamefont {Car},
  \citenamefont {Op~het Veld}, \citenamefont {van Veldhoven}, \citenamefont
  {Koelling}, \citenamefont {Verheijen}, \citenamefont {Pendharkar},
  \citenamefont {Pennachio}, \citenamefont {Shojaei}, \citenamefont {Lee},
  \citenamefont {Palmstr{\o}m}, \citenamefont {Bakkers}, \citenamefont
  {Sarma},\ and\ \citenamefont {Kouwenhoven}}]{Zhang_Nature2018}%
  \BibitemOpen
  \bibfield  {author} {\bibinfo {author} {\bibfnamefont {H.}~\bibnamefont
  {Zhang}}, \bibinfo {author} {\bibfnamefont {C.-X.}\ \bibnamefont {Liu}},
  \bibinfo {author} {\bibfnamefont {S.}~\bibnamefont {Gazibegovic}}, \bibinfo
  {author} {\bibfnamefont {D.}~\bibnamefont {Xu}}, \bibinfo {author}
  {\bibfnamefont {J.~A.}\ \bibnamefont {Logan}}, \bibinfo {author}
  {\bibfnamefont {G.}~\bibnamefont {Wang}}, \bibinfo {author} {\bibfnamefont
  {N.}~\bibnamefont {van Loo}}, \bibinfo {author} {\bibfnamefont {J.~D.~S.}\
  \bibnamefont {Bommer}}, \bibinfo {author} {\bibfnamefont {M.~W.~A.}\
  \bibnamefont {de~Moor}}, \bibinfo {author} {\bibfnamefont {D.}~\bibnamefont
  {Car}}, \bibinfo {author} {\bibfnamefont {R.~L.~M.}\ \bibnamefont {Op~het
  Veld}}, \bibinfo {author} {\bibfnamefont {P.~J.}\ \bibnamefont {van
  Veldhoven}}, \bibinfo {author} {\bibfnamefont {S.}~\bibnamefont {Koelling}},
  \bibinfo {author} {\bibfnamefont {M.~A.}\ \bibnamefont {Verheijen}}, \bibinfo
  {author} {\bibfnamefont {M.}~\bibnamefont {Pendharkar}}, \bibinfo {author}
  {\bibfnamefont {D.~J.}\ \bibnamefont {Pennachio}}, \bibinfo {author}
  {\bibfnamefont {B.}~\bibnamefont {Shojaei}}, \bibinfo {author} {\bibfnamefont
  {J.~S.}\ \bibnamefont {Lee}}, \bibinfo {author} {\bibfnamefont {C.~J.}\
  \bibnamefont {Palmstr{\o}m}}, \bibinfo {author} {\bibfnamefont {E.~P. A.~M.}\
  \bibnamefont {Bakkers}}, \bibinfo {author} {\bibfnamefont {S.~D.}\
  \bibnamefont {Sarma}}, \ and\ \bibinfo {author} {\bibfnamefont {L.~P.}\
  \bibnamefont {Kouwenhoven}},\ }\href {https://doi.org/10.1038/nature26142}
  {\bibfield  {journal} {\bibinfo  {journal} {Nature}\ }\textbf {\bibinfo
  {volume} {556}},\ \bibinfo {pages} {74 EP } (\bibinfo {year}
  {2018})}\BibitemShut {NoStop}%
\bibitem [{\citenamefont {Whiticar}\ \emph {et~al.}()\citenamefont {Whiticar},
  \citenamefont {Fornieri}, \citenamefont {O'Farrell}, \citenamefont
  {Drachmann}, \citenamefont {Wang}, \citenamefont {Thomas}, \citenamefont
  {Gronin}, \citenamefont {Kallaher}, \citenamefont {Gardner}, \citenamefont
  {Manfra}, \citenamefont {Marcus},\ and\ \citenamefont
  {Nichele}}]{Whiticar_2019}%
  \BibitemOpen
  \bibfield  {author} {\bibinfo {author} {\bibfnamefont {A.~M.}\ \bibnamefont
  {Whiticar}}, \bibinfo {author} {\bibfnamefont {A.}~\bibnamefont {Fornieri}},
  \bibinfo {author} {\bibfnamefont {E.~C.~T.}\ \bibnamefont {O'Farrell}},
  \bibinfo {author} {\bibfnamefont {A.~C.~C.}\ \bibnamefont {Drachmann}},
  \bibinfo {author} {\bibfnamefont {T.}~\bibnamefont {Wang}}, \bibinfo {author}
  {\bibfnamefont {C.}~\bibnamefont {Thomas}}, \bibinfo {author} {\bibfnamefont
  {S.}~\bibnamefont {Gronin}}, \bibinfo {author} {\bibfnamefont
  {R.}~\bibnamefont {Kallaher}}, \bibinfo {author} {\bibfnamefont {G.~C.}\
  \bibnamefont {Gardner}}, \bibinfo {author} {\bibfnamefont {M.~J.}\
  \bibnamefont {Manfra}}, \bibinfo {author} {\bibfnamefont {C.~M.}\
  \bibnamefont {Marcus}}, \ and\ \bibinfo {author} {\bibfnamefont
  {F.}~\bibnamefont {Nichele}},\ }\href {https://arxiv.org/abs/1902.07085}
  {\bibinfo  {journal} {arXiv.1902.07085}\ }\BibitemShut {NoStop}%
\bibitem [{\citenamefont {Albrecht}\ \emph {et~al.}(2016)\citenamefont
  {Albrecht}, \citenamefont {Higginbotham}, \citenamefont {Madsen},
  \citenamefont {Kuemmeth}, \citenamefont {Jespersen}, \citenamefont
  {Nyg{\aa}rd}, \citenamefont {Krogstrup},\ and\ \citenamefont
  {Marcus}}]{Albrecht_nature2016}%
  \BibitemOpen
\bibfield  {journal} {  }\bibfield  {author} {\bibinfo {author} {\bibfnamefont
  {S.~M.}\ \bibnamefont {Albrecht}}, \bibinfo {author} {\bibfnamefont {A.~P.}\
  \bibnamefont {Higginbotham}}, \bibinfo {author} {\bibfnamefont
  {M.}~\bibnamefont {Madsen}}, \bibinfo {author} {\bibfnamefont
  {F.}~\bibnamefont {Kuemmeth}}, \bibinfo {author} {\bibfnamefont {T.~S.}\
  \bibnamefont {Jespersen}}, \bibinfo {author} {\bibfnamefont {J.}~\bibnamefont
  {Nyg{\aa}rd}}, \bibinfo {author} {\bibfnamefont {P.}~\bibnamefont
  {Krogstrup}}, \ and\ \bibinfo {author} {\bibfnamefont {C.~M.}\ \bibnamefont
  {Marcus}},\ }\href {https://doi.org/10.1038/nature17162} {\bibfield
  {journal} {\bibinfo  {journal} {Nature}\ }\textbf {\bibinfo {volume} {531}},\
  \bibinfo {pages} {206 EP } (\bibinfo {year} {2016})}\BibitemShut {NoStop}%
\bibitem [{\citenamefont {Deng}\ \emph {et~al.}(2016)\citenamefont {Deng},
  \citenamefont {Vaitiekenas}, \citenamefont {Hansen}, \citenamefont {Danon},
  \citenamefont {Leijnse}, \citenamefont {Flensberg}, \citenamefont {Nyg{\r
  a}rd}, \citenamefont {Krogstrup},\ and\ \citenamefont
  {Marcus}}]{Deng_Science2016}%
  \BibitemOpen
  \bibfield  {author} {\bibinfo {author} {\bibfnamefont {M.~T.}\ \bibnamefont
  {Deng}}, \bibinfo {author} {\bibfnamefont {S.}~\bibnamefont {Vaitiekenas}},
  \bibinfo {author} {\bibfnamefont {E.~B.}\ \bibnamefont {Hansen}}, \bibinfo
  {author} {\bibfnamefont {J.}~\bibnamefont {Danon}}, \bibinfo {author}
  {\bibfnamefont {M.}~\bibnamefont {Leijnse}}, \bibinfo {author} {\bibfnamefont
  {K.}~\bibnamefont {Flensberg}}, \bibinfo {author} {\bibfnamefont
  {J.}~\bibnamefont {Nyg{\r a}rd}}, \bibinfo {author} {\bibfnamefont
  {P.}~\bibnamefont {Krogstrup}}, \ and\ \bibinfo {author} {\bibfnamefont
  {C.~M.}\ \bibnamefont {Marcus}},\ }\href {\doibase 10.1126/science.aaf3961}
  {\bibfield  {journal} {\bibinfo  {journal} {Science}\ }\textbf {\bibinfo
  {volume} {354}},\ \bibinfo {pages} {1557} (\bibinfo {year}
  {2016})}\BibitemShut {NoStop}%
\bibitem [{\citenamefont {Alicea}\ \emph {et~al.}(2011)\citenamefont {Alicea},
  \citenamefont {Oreg}, \citenamefont {Refael}, \citenamefont {von Oppen},\
  and\ \citenamefont {Fisher}}]{Alicea_NatPhys2011}%
  \BibitemOpen
  \bibfield  {author} {\bibinfo {author} {\bibfnamefont {J.}~\bibnamefont
  {Alicea}}, \bibinfo {author} {\bibfnamefont {Y.}~\bibnamefont {Oreg}},
  \bibinfo {author} {\bibfnamefont {G.}~\bibnamefont {Refael}}, \bibinfo
  {author} {\bibfnamefont {F.}~\bibnamefont {von Oppen}}, \ and\ \bibinfo
  {author} {\bibfnamefont {M.~P.~A.}\ \bibnamefont {Fisher}},\ }\href {\doibase
  10.1038/nphys1915} {\bibfield  {journal} {\bibinfo  {journal} {Nature
  Physics}\ }\textbf {\bibinfo {volume} {7}},\ \bibinfo {pages} {412} (\bibinfo
  {year} {2011})}\BibitemShut {NoStop}%
\bibitem [{\citenamefont {Sau}\ \emph {et~al.}(2011)\citenamefont {Sau},
  \citenamefont {Clarke},\ and\ \citenamefont {Tewari}}]{Sau_PRB2011}%
  \BibitemOpen
  \bibfield  {author} {\bibinfo {author} {\bibfnamefont {J.~D.}\ \bibnamefont
  {Sau}}, \bibinfo {author} {\bibfnamefont {D.~J.}\ \bibnamefont {Clarke}}, \
  and\ \bibinfo {author} {\bibfnamefont {S.}~\bibnamefont {Tewari}},\ }\href
  {\doibase 10.1103/PhysRevB.84.094505} {\bibfield  {journal} {\bibinfo
  {journal} {Phys. Rev. B}\ }\textbf {\bibinfo {volume} {84}},\ \bibinfo
  {pages} {094505} (\bibinfo {year} {2011})}\BibitemShut {NoStop}%
\bibitem [{\citenamefont {Hyart}\ \emph {et~al.}(2013)\citenamefont {Hyart},
  \citenamefont {van Heck}, \citenamefont {Fulga}, \citenamefont {Burrello},
  \citenamefont {Akhmerov},\ and\ \citenamefont {Beenakker}}]{Hyart_PRB2013}%
  \BibitemOpen
  \bibfield  {author} {\bibinfo {author} {\bibfnamefont {T.}~\bibnamefont
  {Hyart}}, \bibinfo {author} {\bibfnamefont {B.}~\bibnamefont {van Heck}},
  \bibinfo {author} {\bibfnamefont {I.~C.}\ \bibnamefont {Fulga}}, \bibinfo
  {author} {\bibfnamefont {M.}~\bibnamefont {Burrello}}, \bibinfo {author}
  {\bibfnamefont {A.~R.}\ \bibnamefont {Akhmerov}}, \ and\ \bibinfo {author}
  {\bibfnamefont {C.~W.~J.}\ \bibnamefont {Beenakker}},\ }\href {\doibase
  10.1103/PhysRevB.88.035121} {\bibfield  {journal} {\bibinfo  {journal} {Phys.
  Rev. B}\ }\textbf {\bibinfo {volume} {88}},\ \bibinfo {pages} {035121}
  (\bibinfo {year} {2013})}\BibitemShut {NoStop}%
\bibitem [{\citenamefont {Aasen}\ \emph {et~al.}(2016)\citenamefont {Aasen},
  \citenamefont {Hell}, \citenamefont {Mishmash}, \citenamefont {Higginbotham},
  \citenamefont {Danon}, \citenamefont {Leijnse}, \citenamefont {Jespersen},
  \citenamefont {Folk}, \citenamefont {Marcus}, \citenamefont {Flensberg},\
  and\ \citenamefont {Alicea}}]{Aasen_PRX2016}%
  \BibitemOpen
  \bibfield  {author} {\bibinfo {author} {\bibfnamefont {D.}~\bibnamefont
  {Aasen}}, \bibinfo {author} {\bibfnamefont {M.}~\bibnamefont {Hell}},
  \bibinfo {author} {\bibfnamefont {R.~V.}\ \bibnamefont {Mishmash}}, \bibinfo
  {author} {\bibfnamefont {A.}~\bibnamefont {Higginbotham}}, \bibinfo {author}
  {\bibfnamefont {J.}~\bibnamefont {Danon}}, \bibinfo {author} {\bibfnamefont
  {M.}~\bibnamefont {Leijnse}}, \bibinfo {author} {\bibfnamefont {T.~S.}\
  \bibnamefont {Jespersen}}, \bibinfo {author} {\bibfnamefont {J.~A.}\
  \bibnamefont {Folk}}, \bibinfo {author} {\bibfnamefont {C.~M.}\ \bibnamefont
  {Marcus}}, \bibinfo {author} {\bibfnamefont {K.}~\bibnamefont {Flensberg}}, \
  and\ \bibinfo {author} {\bibfnamefont {J.}~\bibnamefont {Alicea}},\ }\href
  {\doibase 10.1103/PhysRevX.6.031016} {\bibfield  {journal} {\bibinfo
  {journal} {Phys. Rev. X}\ }\textbf {\bibinfo {volume} {6}},\ \bibinfo {pages}
  {031016} (\bibinfo {year} {2016})}\BibitemShut {NoStop}%
\bibitem [{\citenamefont {Flensberg}(2011)}]{Flensberg_PRL2011}%
  \BibitemOpen
  \bibfield  {author} {\bibinfo {author} {\bibfnamefont {K.}~\bibnamefont
  {Flensberg}},\ }\href {\doibase 10.1103/PhysRevLett.106.090503} {\bibfield
  {journal} {\bibinfo  {journal} {Phys. Rev. Lett.}\ }\textbf {\bibinfo
  {volume} {106}},\ \bibinfo {pages} {090503} (\bibinfo {year}
  {2011})}\BibitemShut {NoStop}%
\bibitem [{\citenamefont {Kitaev}(2001)}]{Kitaev_2001}%
  \BibitemOpen
  \bibfield  {author} {\bibinfo {author} {\bibfnamefont {A.~Y.}\ \bibnamefont
  {Kitaev}},\ }\href {\doibase 10.1070/1063-7869/44/10s/s29} {\bibfield
  {journal} {\bibinfo  {journal} {Physics-Uspekhi}\ }\textbf {\bibinfo {volume}
  {44}},\ \bibinfo {pages} {131} (\bibinfo {year} {2001})}\BibitemShut
  {NoStop}%
\bibitem [{\citenamefont {Kamenev}(2011)}]{Kamenev_2011}%
  \BibitemOpen
  \bibfield  {author} {\bibinfo {author} {\bibfnamefont {A.}~\bibnamefont
  {Kamenev}},\ }\href {\doibase 10.1017/CBO9781139003667} {\emph {\bibinfo
  {title} {Field Theory of Non-Equilibrium Systems}}}\ (\bibinfo  {publisher}
  {Cambridge University Press},\ \bibinfo {year} {2011})\BibitemShut {NoStop}%
\bibitem [{\citenamefont {Souto}\ \emph {et~al.}(2016)\citenamefont {Souto},
  \citenamefont {Mart\'{\i}n-Rodero},\ and\ \citenamefont
  {Yeyati}}]{Souto_PRL2016}%
  \BibitemOpen
  \bibfield  {author} {\bibinfo {author} {\bibfnamefont {R.~S.}\ \bibnamefont
  {Souto}}, \bibinfo {author} {\bibfnamefont {A.}~\bibnamefont
  {Mart\'{\i}n-Rodero}}, \ and\ \bibinfo {author} {\bibfnamefont {A.~L.}\
  \bibnamefont {Yeyati}},\ }\href {\doibase 10.1103/PhysRevLett.117.267701}
  {\bibfield  {journal} {\bibinfo  {journal} {Phys. Rev. Lett.}\ }\textbf
  {\bibinfo {volume} {117}},\ \bibinfo {pages} {267701} (\bibinfo {year}
  {2016})}\BibitemShut {NoStop}%
\bibitem [{\citenamefont {Souto}\ \emph {et~al.}(2017)\citenamefont {Souto},
  \citenamefont {Mart\'{\i}n-Rodero},\ and\ \citenamefont
  {Yeyati}}]{Souto_PRB2017}%
  \BibitemOpen
  \bibfield  {author} {\bibinfo {author} {\bibfnamefont {R.~S.}\ \bibnamefont
  {Souto}}, \bibinfo {author} {\bibfnamefont {A.}~\bibnamefont
  {Mart\'{\i}n-Rodero}}, \ and\ \bibinfo {author} {\bibfnamefont {A.~L.}\
  \bibnamefont {Yeyati}},\ }\href {\doibase 10.1103/PhysRevB.96.165444}
  {\bibfield  {journal} {\bibinfo  {journal} {Phys. Rev. B}\ }\textbf {\bibinfo
  {volume} {96}},\ \bibinfo {pages} {165444} (\bibinfo {year}
  {2017})}\BibitemShut {NoStop}%
\bibitem [{SM()}]{SM}%
  \BibitemOpen
  \href@noop {} {\bibinfo  {journal} {See Supplemental Material for details on
  the numerical methods used in this work, which includes Refs.
  \cite{Martin_Advances,Zazunov_PRB2016,Bondyopadhaya_PRB2019}}\ }\BibitemShut
  {NoStop}%
\bibitem [{\citenamefont {Higginbotham}\ \emph {et~al.}(2015)\citenamefont
  {Higginbotham}, \citenamefont {Albrecht}, \citenamefont {Kirsanskas},
  \citenamefont {Chang}, \citenamefont {Kuemmeth}, \citenamefont {Krogstrup},
  \citenamefont {Jespersen}, \citenamefont {Nyg{\aa}rd}, \citenamefont
  {Flensberg},\ and\ \citenamefont {Marcus}}]{Higginbotham_NatPhys2015}%
  \BibitemOpen
\bibfield  {journal} {  }\bibfield  {author} {\bibinfo {author} {\bibfnamefont
  {A.~P.}\ \bibnamefont {Higginbotham}}, \bibinfo {author} {\bibfnamefont
  {S.~M.}\ \bibnamefont {Albrecht}}, \bibinfo {author} {\bibfnamefont
  {G.}~\bibnamefont {Kirsanskas}}, \bibinfo {author} {\bibfnamefont
  {W.}~\bibnamefont {Chang}}, \bibinfo {author} {\bibfnamefont
  {F.}~\bibnamefont {Kuemmeth}}, \bibinfo {author} {\bibfnamefont
  {P.}~\bibnamefont {Krogstrup}}, \bibinfo {author} {\bibfnamefont {T.~S.}\
  \bibnamefont {Jespersen}}, \bibinfo {author} {\bibfnamefont {J.}~\bibnamefont
  {Nyg{\aa}rd}}, \bibinfo {author} {\bibfnamefont {K.}~\bibnamefont
  {Flensberg}}, \ and\ \bibinfo {author} {\bibfnamefont {C.~.~M.}\ \bibnamefont
  {Marcus}},\ }\href {https://doi.org/10.1038/nphys3461} {\bibfield  {journal}
  {\bibinfo  {journal} {Nature Physics}\ }\textbf {\bibinfo {volume} {11}},\
  \bibinfo {pages} {1017 EP } (\bibinfo {year} {2015})}\BibitemShut {NoStop}%
\bibitem [{\citenamefont {Mart\'in-Rodero}\ and\ \citenamefont
  {Yeyati}(2011)}]{Martin_Advances}%
  \BibitemOpen
  \bibfield  {author} {\bibinfo {author} {\bibfnamefont {A.}~\bibnamefont
  {Mart\'in-Rodero}}\ and\ \bibinfo {author} {\bibfnamefont {A.~L.}\
  \bibnamefont {Yeyati}},\ }\href {\doibase 10.1080/00018732.2011.624266}
  {\bibfield  {journal} {\bibinfo  {journal} {Advances in Physics}\ }\textbf
  {\bibinfo {volume} {60}},\ \bibinfo {pages} {899} (\bibinfo {year}
  {2011})}\BibitemShut {NoStop}%
\bibitem [{\citenamefont {Zazunov}\ \emph {et~al.}(2016)\citenamefont
  {Zazunov}, \citenamefont {Egger},\ and\ \citenamefont
  {Levy~Yeyati}}]{Zazunov_PRB2016}%
  \BibitemOpen
  \bibfield  {author} {\bibinfo {author} {\bibfnamefont {A.}~\bibnamefont
  {Zazunov}}, \bibinfo {author} {\bibfnamefont {R.}~\bibnamefont {Egger}}, \
  and\ \bibinfo {author} {\bibfnamefont {A.}~\bibnamefont {Levy~Yeyati}},\
  }\href {\doibase 10.1103/PhysRevB.94.014502} {\bibfield  {journal} {\bibinfo
  {journal} {Phys. Rev. B}\ }\textbf {\bibinfo {volume} {94}},\ \bibinfo
  {pages} {014502} (\bibinfo {year} {2016})}\BibitemShut {NoStop}%
\bibitem [{\citenamefont {Bondyopadhaya}\ and\ \citenamefont
  {Roy}(2019)}]{Bondyopadhaya_PRB2019}%
  \BibitemOpen
  \bibfield  {author} {\bibinfo {author} {\bibfnamefont {N.}~\bibnamefont
  {Bondyopadhaya}}\ and\ \bibinfo {author} {\bibfnamefont {D.}~\bibnamefont
  {Roy}},\ }\href {\doibase 10.1103/PhysRevB.99.214514} {\bibfield  {journal}
  {\bibinfo  {journal} {Phys. Rev. B}\ }\textbf {\bibinfo {volume} {99}},\
  \bibinfo {pages} {214514} (\bibinfo {year} {2019})}\BibitemShut {NoStop}%
\end{thebibliography}%

\newpage
\widetext
\appendix

\section*{Supplementary information on ``Time scales for charge transfer based operations on Majorana systems"}
\section{Green functions formalism}
\label{sec::formalism}
\begin{figure}
     \begin{minipage}{0.8\linewidth}
      \includegraphics[width=0.6\textwidth]{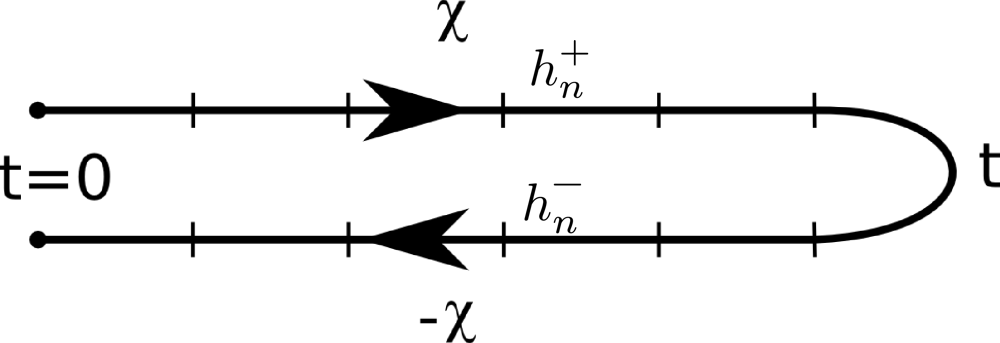}
     \end{minipage}
\caption{Keldysh contour considered to analyze the transient regime. $\chi$ indicates the counting field changing sign on the two branches of the contour and $\delta t$ corresponds to the time step in the discretized calculation of the generating function $Z(\chi,t)$.}
\label{contour}
\end{figure}

In this section of the supplementary information (SI) the main theoretical tools are introduced, summarizing the most important expressions describing the system time evolution. We employ a Green functions formalism on the Keldysh-Nambu contour. In this framework, the inverse QD Green function is given by Eq. (3) of the main text. In order to describe the time dynamics, the Keldysh contour is discretized in time with a time step, $\delta t$, smaller than any other inverse energy (see Fig. \ref{contour}). The expression for the bare inverse QD Green function on the discretized Keldysh contour is given by
\begin{equation}
i {\bf g}^{-1}_{e} = \left(\begin{array}{cccc|cccc} -1 & & & & & & & -\rho \\
h_{1}^- & -1 & & & & & &  \\
& \ddots & \ddots & & & & & \\
& & h_{N-1}^- & -1 & &  & &  \\
\hline  
&  & & 1 & -1 & & &  \\
&  & & & h_{N-1}^+ & -1 & &   \\
&  &  &  & & \ddots & \ddots &  \\
&  &  & & & & h_{1}^+ & -1 \end{array} \right)_{2N\times2N}\; ,
\label{kamenev}
\end{equation}
where $h^{\pm}_n(t) = 1 \mp i\delta t[\epsilon_0(t)+\epsilon_0(t+\delta t)]/2$ and the matrix dimension is given by $N=t/\delta t$, with $t$ the final time. The boldphase letter is used to denote Green functions on the discretized Keldysh contour. In Eq. (\ref{kamenev}) the initial QD charge, $n_e$, which enters the expression through $\rho=n_e(0)/[1-n_e(0)]$, is set either to $n_e(0)=0$ or $1$, depending on the sign of the initial level position ($|\epsilon_i|\gg\Delta$). The Nambu structure of the inverse QD Green function can be written as
\begin{equation}
	{\bf g}^{-1}_0 = \left( \begin{array}{cc} g^{-1}_e & 0 \\ 0 & g^{-1}_h \end{array} \right), 
\end{equation}
where the hole part is given by $g^{-1}_h= g^{-1}_e[-\epsilon_0(t),(1-n_e(0))]$.

The tunneling self-energy of Eq. (3) of the main text describes the coupling to the TS electrodes, which are considered to be infinitely long Kitaev chains in the topological regime, described by the Hamiltonian
\begin{eqnarray}
H_{\nu}=-t_0\sum_n(c_{n+1,\nu}^\dagger c_{n,\nu}-\mbox{H.C.})-
\mu\sum_n c_{n,\nu}^\dagger c_{n,\nu}+\Delta(c_{n,\nu}^\dagger c_{n+1,\nu}+\mbox{H.C.})\,,
\label{KitaevChain}
\end{eqnarray}
where $c_{n,\nu}$ is the annihilation operator acting on site $n$ of electrode $\nu$. In Eq. (\ref{KitaevChain}) $t_0$ is the hopping inside the chain, $\mu$ is the chemical potential, taken to zero for simplicity, and $\Delta$ is the superconducting gap. For simplicity, $\Delta$ will be taken real and positive, and the superconducting phase factors are introduced in the superconducting phase factors introduced in the tunneling Hamiltoninan. The retarded/advanced (r/a) components of the electrodes Green function are given in the wideband limit by (derivation details can be found in Ref. \cite{Zazunov_PRB2016})
\begin{equation}
	\hat{g}_{L,R}^{r}(\omega)=\frac{2}{|t_0|}\frac{\sqrt{\Delta^2-(\omega+ i\eta)^2}\hat{\tau}_0\pm\Delta\hat{\tau}_x}{(\omega+ i\eta)}\,,
\end{equation}
with $\eta$ being a positive infinitesimal and the Pauli matrices are acting in the Nambu space, represented by $\hat{ }$. The $\pm$ sign originates from the fact that the QD is attached to the rightmost (leftmost) site of the left (right) TS (see Fig. 1 b of the main text). Using general relations in the equilibrium situation \cite{Martin_Advances}, the Keldysh off-diagonal components of the self-energy in time domain are given by
\begin{equation}
	\hat{g}^{+-}_{\nu}(t,t')=-\int d\omega \left[\hat{g}_{\nu}^{a}(\omega)-\hat{g}_{\nu}^{r}(\omega)\right]n_F(\omega)e^{-i\omega (t-t')}\,,
	\label{self-energy}
\end{equation}
where $\hat{g}_{\nu}^{a}=(\hat{g}_{\nu}^{r})^*$ and $n_F$ is the Fermi function. A similar expression can be obtained for the $-+$ components, by replacing $n_F$ by $n_F-1$. The self-energy can be obtained from the lead Green functions as 
\begin{equation}
	\hat{\Sigma}^{+-}_\nu(t,t')=\theta(t)\theta(t')|V_\nu|^2 e^{-i\hat{\tau}_z\phi_\nu/2} \hat{g}^{+-}_{\nu}(t,t')e^{i\hat{\tau}_z\phi_\nu/2}\,.
\end{equation}
In order to integrate Eq. (\ref{self-energy}) we use the property $1/(\omega\pm i\eta)=\mp i\pi\delta(\omega)+\mbox{PV}(1/\omega)$, which divides the self-energy into two contributions. The first one, involving the Dirac-delta functional, describes the tunneling between the QD and the MBSs ($\Sigma_{MBS}$). The second part, which is related to the integral of the principal value, contains information about the continuum degrees of freedom ($\Sigma_{c}$). The total self-energy is then given by
\begin{equation}
	\hat{\Sigma}(t,t')=\sum_\nu\left[\hat{\Sigma}_{MBS;\nu}(t,t')+\hat{\Sigma}_{c;\nu}(t,t')\right]\,.
	\label{total_self-energy}
\end{equation}
The contribution from the MBSs can be integrated exactly, giving
\begin{equation}
	\hat{\Sigma}_{MBS;\nu}^{+-}(t,t')=-\hat{\Sigma}_{MBS;\nu}^{-+}(t,t')=-i\Gamma_\nu(\hat{\tau}_0\pm\hat{\tau}_x)\,,
	\label{MBS_self-energy}
\end{equation}
where $\Gamma_\nu=\pi |V_\nu|^2 \rho_F$ and $\rho_F=2/\pi |t_0|$ is the normal density of states at the Fermi level. The contribution from the continuum comes from the Fourier transform of
\begin{equation}
	\hat{\Sigma}_{c;\nu}^{r/a}(\omega)=\Gamma_{\nu}\,e^{-i\hat{\tau}_z\phi_{\nu}/2}\frac{\sqrt{\Delta^2-\omega^2}}{\omega}\hat{\tau}_0e^{i\hat{\tau}_z\phi_{\nu}/2}\,,
\end{equation}
which describes the electronic states outside the superconducting gap. 

Finally, the Keldysh diagonal components are given by general relations
\begin{eqnarray}
	\hat{\Sigma}^{++}(t,t')&=&-\theta(t,t')\hat{\Sigma}^{-+}(t,t')-\theta(t'-t)\hat{\Sigma}^{+-}(t,t')\,,\nonumber\\
	\hat{\Sigma}^{--}(t,t')&=&-\theta(t,t')\hat{\Sigma}^{+-}(t,t')-\theta(t'-t)\hat{\Sigma}^{-+}(t,t')\,.
\end{eqnarray}

Using the self-energies derived above, Eq. (3) from the main text can be numerically inverted, which provides access to the mean QD properties, such as the time evolution of the population as $\left\langle n_d(t)\right\rangle =\mbox{Im}[G(t,t)]^{+-}_{11}$, where the subindices indicate the Nambu component. The average parity of each lead is given by $\left\langle p_\nu\right\rangle=i\left\langle \gamma_{\nu}^A\gamma_{\nu}^B\right\rangle$. For non-overlapping MBSs it can be determined as 
\begin{equation}
	\left\langle p_\nu(t)\right\rangle=\left\langle p_\nu(0)\right\rangle\left\{1-2i\Gamma_\nu \sum_{j,k=1,2}s_{\nu}^{j+k} e^{i\hat{\tau}_z\phi_\nu/2}\int_{0}^{t}dt_1\int_{0}^{t}dt_2 \left[G^{+-}_{jk}(t_1,t_2)-G^{-+}_{jk}(t_1,t_2)\right]e^{-i\hat{\tau}_z\phi_\nu/2}\right\}\,,
	\label{meanPop}
\end{equation}
where the sum is performed in the Nambu space and $s_{L}=1$ and $s_R=-1$.

\section{Full counting statistics}
\label{sec::FCS}
The full counting statistics (FCS) method, which analyzes the statistics of the transferred electrons, is an ideal tool for extracting the probability of successful manipulation. This information is absent in the QD Green functions, given in Eq. (3) of the main text. While FCS has traditionally been used to investigate the charge and current fluctuations of nanojunctions, in this section we extend this formalism for addressing the parity change of a topological system. The number of electrons transferred are determined using a counting variable $\chi$, which enters as phase factors in the tunneling amplitudes, changing sign in the Keldysh contour as indicated in Fig. \ref{contour}. The main ingredient in the theory is the so-called generating function (GF)
\begin{equation}
	Z(\chi,t)=\sum_n \mathcal{P}_{n}(t)e^{i\,n\,\chi}\,,
	\label{GF_def}
\end{equation}
 which encodes the probability distribution of transferred charges until time $t$, $\mathcal{P}_n(t)$. 
 An interesting particular case corresponds to taking $\chi=\pi$. In this particular case, the GF is given by $Z(\chi,t)=\sum_n \mathcal{P}_{n}(t)(-1)^{n}$, which describes the probability of a parity change (transfer of an odd number of electrons).

To analyze the parity change of the subgap states, we will include the counting variable only in the MBS self-energies (\ref{MBS_self-energy}) as
\begin{equation}
	\hat{\bar{\Sigma}}_{M;\nu}(\chi)=e^{i\tau^{K}_z(\hat{\tau}_z+\hat{\tau}_0)\chi_\nu/2}\Sigma_{M;\nu}e^{i\tau^{K}_z(\hat{\tau}_z+\hat{\tau}_0)\chi_\nu/2}\,,
	\label{MBS_chi}
\end{equation}
where $\tau^{K}_z$ is the Pauli matrix acting in the Keldysh space and $\chi_\nu$ the counting field variable for each electrode. The counting field in Eq.  (\ref{MBS_chi}) only projects the first Nambu component, which is sufficient to determine the system state. There are two different choices for the counting field at both sides of the junction. The first possibility, which is used in the main text, corresponds to taking $\chi_L=\chi_R=\chi/2$. This situation corresponds to projecting the system wavefunction onto a state with well defined joint parity of left and right MBSs. Alternatively, the counting field variable can be set to zero on on side and to $\chi$ on the other one. This situation corresponds to the projection of the state of one TS onto a state with well defined parity for $\chi=\pi$. In this case, however, the charge-transfer operation is affected by the measurement process, since an instantaneous projection of the state on one side will set the MBS parity either to even or odd, avoiding the formation of quantum superpositions. As an example, this choice will prevent the system from returning back to its original state after the protocol described in Fig. 4 of the main text. 

Independently from the choice of counting field, the GF in the discrete time mesh can be computed as 
\begin{equation}
	Z(\chi,t)=\frac{\mbox{det}\left[{\bf g}_{0}^{-1}-{\bf \Sigma(\chi)}\right]}{\mbox{det}\left[{\bf g}_{0}^{-1}-{\bf \Sigma}(\chi=0)\right]}\,.
\end{equation} 
Finally, it is worth commenting that the same expression can be used to compute the transferred charge  and current cumulants by taking successive derivatives of the GF as $\left\langle c_k(t)\right\rangle=\partial^k/\partial (i\chi)^k \,\mbox{Log}[Z(\chi,t)]_{\chi=0}$ and $\left\langle I^k(t)\right\rangle=\partial/\partial t \,\left\langle c_k(t)\right\rangle$, for $\chi_L=-\chi_R=\chi/2$.

\section{MBS parity oscillations}
\label{sec::dephasing}
The charge-transfer based operations are not protected by topology. Therefore, they can suffer from decoherence and processes affecting the MBSs parity. This is the case for the parity oscillations shown in the right panel of Fig. 2 of the main text. They occur for any superconducting phase difference, except for $\phi=\pi$ (mod $2\pi$), where the final state after an adiabatic operation  at zero temperature does not depend on how it is performed. In this section of the SI we analyze the absence of parity oscillations at this particular phase, discussing the effect of small deviations from $\phi=\pi$ (mod $2\pi$). In the following we will consider an adiabatic operation and focus on the zero temperature case, i.e. no excited quasiparticles above the superconducting gap.

We consider the effective Hamiltonian
\begin{equation}
 H_{eff}=H_{QD}+H_{c}+H_{T}\,,
 \label{H_eff}
\end{equation}
where the Hamiltonian of the continuum of states above the superconducting gap is given by
\begin{equation}
 H_c=\sum_{k,\nu=L,R}\left[\xi_k c_{k,\nu}^{\dagger}c_{k,\nu}-\Delta\left( c_{k,\nu}c_{-k,\nu}+\mbox{H.C.}\right)\right]\,.
\end{equation}
This term can be diagonalized using the Bogoliuvob transformation, finding
\begin{equation}
 \tilde{H}_c=\sum_{k,\nu=L,R}E_k\gamma_{k,\nu}^{\dagger}\gamma_{k,\nu}\,,
\end{equation}
where $E_k=\sqrt{\xi_{k}^2+\Delta^2}$ and the tilde is used to denote the transformed Hamiltonian. The Bogoliuvob quasiparticle operators are given by $\gamma_{-k,\nu}^{\dagger}=u_{k,\nu}^*c_{-k,\nu}^\dagger+v_{k,\nu}^*c_{k,\nu}$ and $\gamma_{k,\nu}=u_{k,\nu}c_{k,\nu}-v_{k,\nu}c_{-k,\nu}^\dagger$, with $|u_{k}^2|=(1-\xi_k/E_k)/2$ and $|v_{k}^2|=(1+\xi_k/E_k)/2$.

The Hamiltonian describing electron tunneling between the QD and the TS electrodes can be written as a sum of two contributions $\tilde{H}_T=\tilde{H}_{T,MBS}+\tilde{H}_{T,c}$, describing the tunneling to the MBSs and the continuum of states, respectively. These two contributions are given by
\begin{equation}
 \tilde{H}_{T,MBS}=V_L\gamma_{L}^A(e^{i\phi_L/2}d-e^{-i\phi_L/2}d^\dagger)+V_R\gamma_{R}^B(e^{i\phi_R/2}d-e^{-i\phi_R/2}d^\dagger)\,,
\label{H_TMBS}
\end{equation}
where $\gamma_{L}^A=c_{L}+c_{L}^\dagger$ and $\gamma_{R}^B=i(c_{R}-c_{R}^\dagger)$ are the MBS quasiparticle operators and
\begin{equation}
\tilde{H}_{T,c}=\sum_{k,\nu}V_{k,\nu}\left[e^{-i\phi_\nu/2}d^\dagger \left(u^{*}_{k,\nu}\gamma_{k,\nu}+v_{k,\nu}\gamma_{k,\nu}^\dagger\right)+\mbox{H.C.}\right]\,.
\end{equation}
The time evolution of a given initial state in the interaction picture can be obtained through
\begin{equation}
 \left| \Psi(t)\right\rangle=\mathcal{T} e^{\int_{0}^t \tilde{H}_T(t')dt'}\left| \Psi(0)\right\rangle=\mathcal{T} \left[1+\int_{0}^t dt_1 \tilde{H}_T(t_1)+\frac{1}{2}\int_{0}^t\int_{0}^t dt_1dt_2 \tilde{H}_T(t_1)\tilde{H}_T(t_2)+\hdots\right]\left| \Psi(0)\right\rangle
 \label{state_evol}
\end{equation}
where $\mathcal{T}$ is the time ordering operator. The different terms in the right hand side of Eq. (\ref{state_evol}) describe all the possible tunneling events between the QD and the TSs. The linear term in $\tilde{H}_T$ describes the process in which the QD charge state is transferred to the TSs during the charge-transfer based operation, while higher order terms describe undesired effects such as the MBS parity oscillations shown in Fig. 2 of the main text. At second order in Eq. (\ref{state_evol}), the only terms affecting the MBS parities are the ones involving the transfer of an electron between the left and right MBSs. In the expansion, two of these terms leading to the same final state are given by
\begin{equation}
-\mathcal{T}\left\{V_L V_R \int dt_1 \int dt_2 \left[e^{i(\phi_R-\phi_L)/2}\gamma^{A}_L(t_1) d^\dagger(t_1) \gamma^{B}_{R}(t_2) d(t_2)+e^{i(\phi_L-\phi_R)/2}\gamma^{B}_{R}(t_1) d^\dagger(t_1) \gamma^{A}_L(t_2) d(t_2)\right]\right\}\,.
\label{H_T_2_MBS}
\end{equation}
These processes give as a consequence an electron transferred from the left to the right TS, changing the MBS parity at both sides of the junction. For a phase difference $\phi=\pi$ (mod $2\pi$), the two contributions in Eq. (\ref{H_T_2_MBS}) cancel each other, negating the undesired parity changes between the MBSs. A similar cancellation can be found for the remaining second order terms affecting the MBSs parity state. In a similar way, for any higher order term in $\tilde{H}_{T,MBS}$ in Eq. (\ref{state_evol}), there is always another one of the same order cancelling its contribution for  $\phi=\pi$ (mod $2\pi$).

At zero temperature, the continuum of states outside the superconducting gap start to contribute to the MBS parity at fourth order in $\tilde{H}_T$ in Eq. (\ref{state_evol}). It describes processes where electrons (or holes) coming from the continuum of states can tunnel to the MBSs, mediated by the QD. Two of these processes leading to the same final state are described by
\begin{eqnarray}
\mathcal{T}\left\{ V_L V^{3}_R\int dt_1 \hdots\int dt_4 \left[e^{i(\phi_R-\phi_L)/2}\tilde{H}_{T,c}(t_4)  \gamma^{A}_L(t_3)d^\dagger(t_3) \gamma^{B}_{R}(t_2) d(t_2) \tilde{H}_{T,c}(t_1)+\right.\right.\nonumber\\
\left.\left.e^{i(\phi_R-\phi_L)/2}\tilde{H}_{T,c}(t_4) \gamma^{B}_{R}(t_3) d^\dagger(t_3) \gamma^{A}_L(t_2)d(t_2) \tilde{H}_{T,c}(t_1)\right]\right\}\,,
\end{eqnarray}
which cancel each other for $\phi=\pi$ (mod $2\pi$). It can be demonstrated that, for any other higher order term affecting the MBSs parity state, there exists another one of the same order cancelling it for this phase difference. It implies that the states above the superconducting gap do not induce changes on the MBSs parity state at $T=0$ for $\phi=\pi$ (mod $2\pi$). Therefore, the main errors occurring during the operation for these parameters are due to non-adiabatic effects. We note that dephasing is not important in the analyzed parameter regime, as shown in Fig. 4.

For $\phi \neq (2n+1)\pi$, the higher order contributions in $\tilde{H}_T$ in Eq. (\ref{state_evol}) do not cancel. They give rise to a time oscillation of the MBSs parity as shown in Fig. 2 of the main text, leading to an oscillating current \cite{Bondyopadhaya_PRB2019} in time, due to the breaking of degeneracy between the even and odd parity states of the non-local fermion formed by $\gamma^{A}_L$ and $\gamma^{B}_R$. Because of these oscillations, the final state depends on the details of the operation (such as sweep rate), even in the adiabatic limit.  Also, they can lead to a result similar to the fully dephased situation  when the QD energy is not far enough above the superconducting gap, if the characteristic measurement time on MBS parity is larger than the oscillation period.

Slightly away from the ideal condition, $\phi=(2n+1)\pi+\delta\phi$, the higher order terms in the expansion of Eq. (\ref{state_evol}) start to contribute. In order to minimize their contribution, the manipulation time should be faster than their characteristic time scales given by $\sqrt{\epsilon^{2}_0+4(V^{2}_L+V^{2}_R)}/(4V_L V_R \delta\phi)$ for $\delta\phi\ll1$. It defines an upper limit for the operation time scale, illustrating the importance of accurately control the phase difference between the TSs. 

Finally, it is worth commenting that this demonstration is based on several assumptions. Firstly, we have assumed that the electrons tunneling from a given lead have a well-defined phase. Additionally, the demonstration is based on the assumption that
adding and removing electrons from the MBSs lead to the same final state, as there is no charging energy in the TSs. Therefore, it is important that superconductors are grounded. Furthermore, a coupling between the MBSs in the same TS, lifting the even-odd ground state degeneracy, can be detrimental for the charge-based operations, as shown in the next section using a simplified version of the effective Hamiltonian (\ref{H_eff}). Finally, a non-zero temperature or a non-adiabatic operation can induce errors during the QD manipulation, as discussed in the main text.

\section{Low energy model: overlap between MBSs.}
\renewcommand{\thefigure}{C.\arabic{figure}}
\label{sec::LEM}

In this section, we introduce a simplified model that includes only the low energy degrees of freedom. It disregards the electronic degrees of freedom outside the superconducting gap. The low energy Hamiltonian is given by
\begin{equation}
	H_{LEM}=H_{QD}+H_{T,MBS}+H_{o,MBS}\,,
	\label{H_eff}
\end{equation}
where $H_{QD}=\epsilon_0(t) d^\dagger d$ with $\epsilon_0(t)$ the time-dependent energy level and the tunneling between the MBSs and the dot is described by Eq. (\ref{H_TMBS}). Finally, $H_{o,MBS}=\sum_{\nu=L,R}\eta_\nu\gamma_{\nu}^A\gamma_{\nu}^B$ describes the overlap between MBSs at the same wire.

Using the low energy Hamiltonian of Eq. (\ref{H_eff}), the time evolution of a given initial state, $\left|\Psi(0)\right\rangle$, is described by unitary evolution as
\begin{equation}
	\left|\Psi(t+\delta t)\right\rangle=\exp\left(-iH_{LEM}\delta t\right)\left|\Psi(t)\right\rangle\,,
\end{equation}
where $\delta t$ has to be taken much smaller than the smallest inverse energy in the model to ensure convergence. The simplified model provided in Eq. (\ref{H_eff}) correctly describes the time evolution of the system for adiabatic manipulations, as shown in the top panel of Fig. 4 of the main text. However, it does not describe the effect of thermally excited quasiparticles in the TSs and the relaxation of QD charge to the continuum of states for non-adiabatic manipulations.

In agreement with the full calculation, for $V_L=V_R$, $\phi_L-\phi_R=2n\pi$ and $\eta_\nu\neq0$, the Hamiltonian of Eq.  (\ref{H_eff}) describes the existence of dark states partially decoupling subspaces with different total MBS parity states. They are given by $(\left|1,1\right\rangle-i\left|-1,-1\right\rangle)/\sqrt{2}$ and $(\left|1,-1\right\rangle-i\left|-1,1\right\rangle)/\sqrt{2}$ for the even and odd MBS parities respectively. These states are decoupled from each other, limiting the success probability to $P=1/2$ for an initial state with a well defined MBS parity at both sides of the junction. In a more general situation with $\phi_L-\phi_R\neq 2n\pi$, the probability can reach values close to unity. However, the manipulation time depends strongly on the superconducting phase difference, as shown in the right panel of Fig. 2 of the main text.

\begin{figure}
	\includegraphics[width=0.6\textwidth]{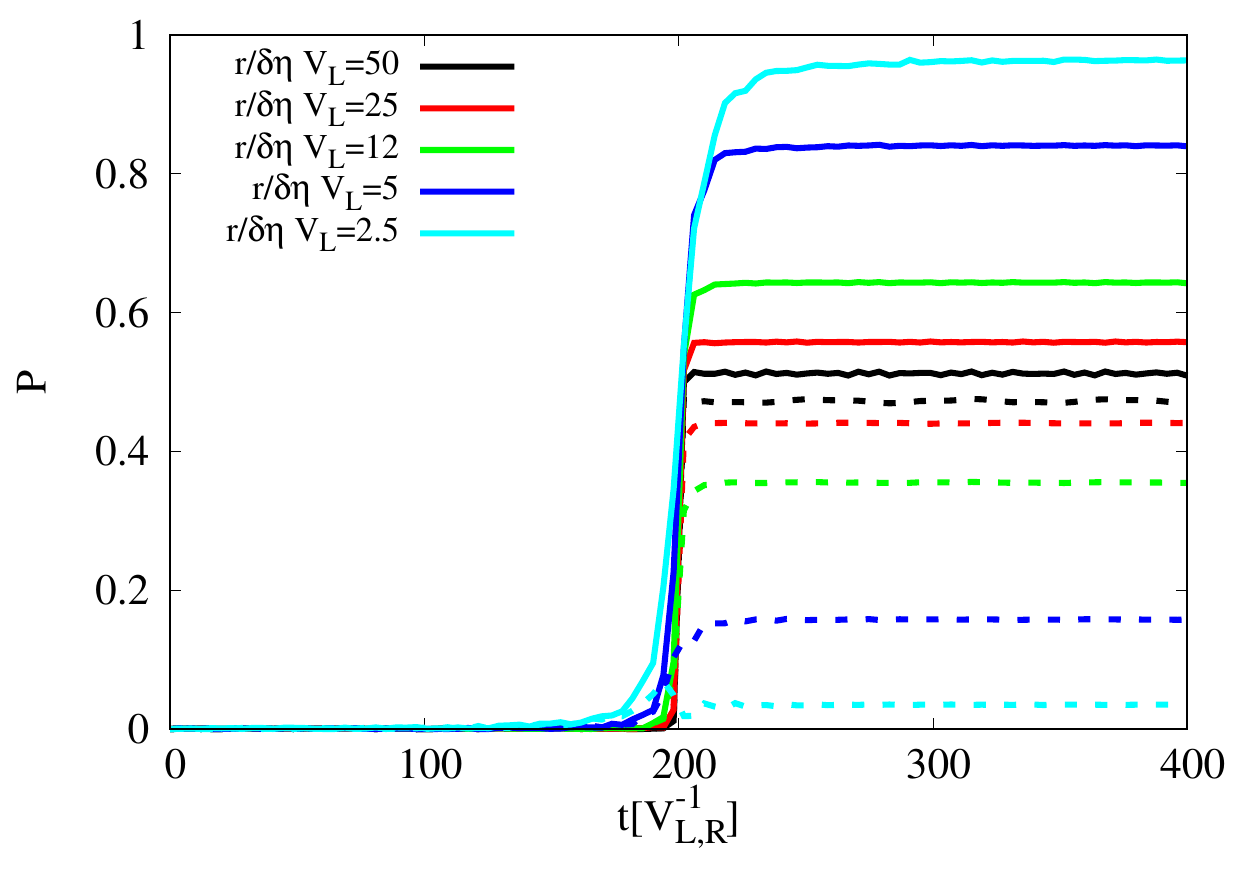}
	\caption{Occupation of the global ground state (solid lines) and excited state (dashed lines) of the same parity after a manipulation. Results are showed for an asymmetric coupling between MBSs in the left and right lead, $\eta_L=V_L/10$, $\eta_R=\eta_L/2$ and $\delta \eta=|\eta_L-\eta_R|$. The remaining parameters are $V_L=V_R$ and $\phi=\pi$.}
	\label{coupled_MBS}
\end{figure}

The model in Eq. (\ref{H_eff}) allows us to easily incorporate the coupling between MBSs in the same wire, which lifts their ground state degeneracy. For $\eta_L=\eta_R$ and starting from the ground state of the two TSs, an adiabatic QD level manipulation also leads to a coherent superposition between even and odd MBS parity states for each electrode. In contrast, for $\eta_L\neq\eta_R$ an adiabatic manipulation leads to a convergence to the global ground state of the system, as illustrated by the blue lines in Fig. \ref{coupled_MBS}. Therefore, $1/\delta\eta$ defines a lower bound for the manipulation rate, with $\delta\eta=|\eta_L-\eta_R|$.

\begin{figure}
	\includegraphics[width=0.6\textwidth]{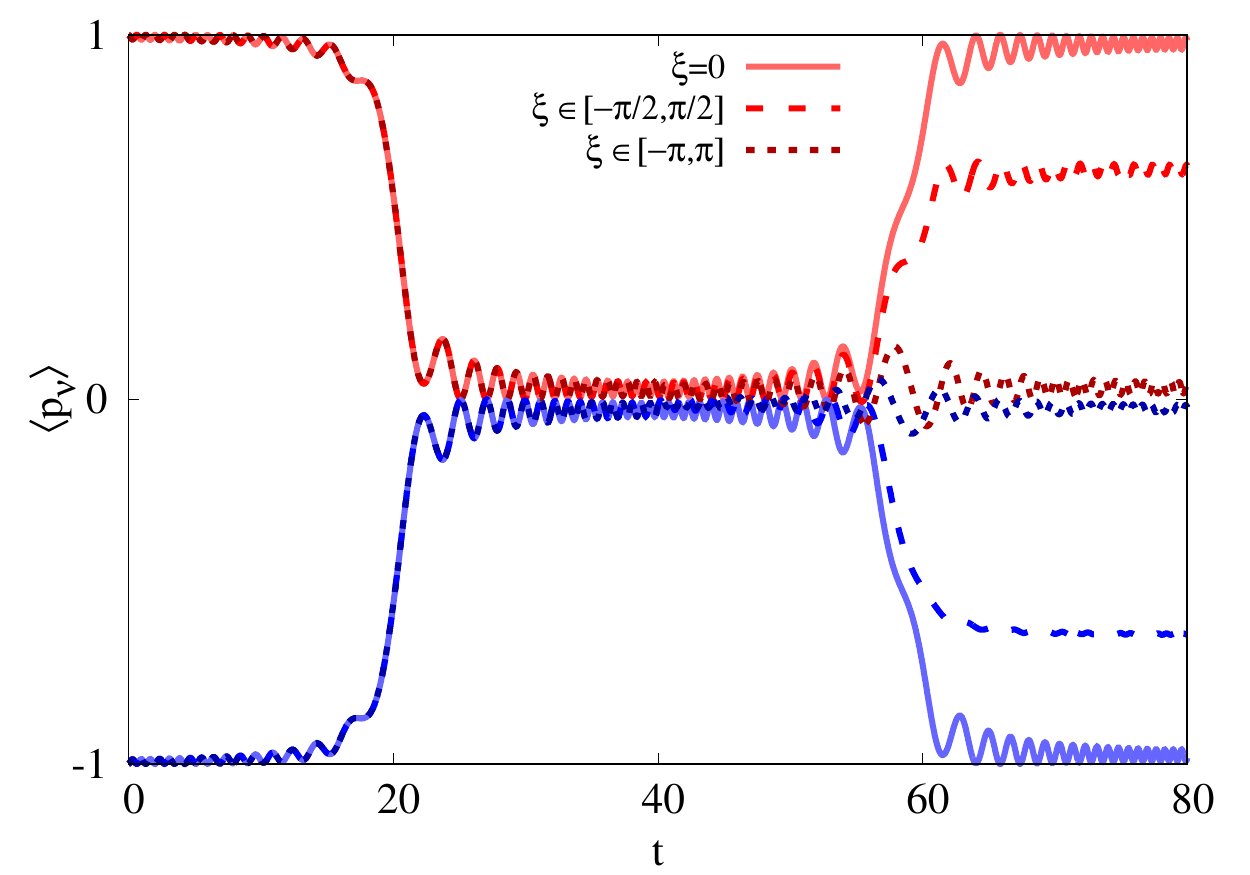}
	\caption{Parity of the left (red curves) and right (blue curves) MBSs for different values of a random phase introduced in the quantum state after the first manipulation (see text for details). The parameters are $V_L=V_R=0.39$, $\phi=\pi$ and $r=1/2$. Time units have chosen to be the asme as in Fig. 4 of the main text.}
	\label{decoherence}
\end{figure}

Another question of interest that can be analyzed using the model of Eq. (\ref{H_eff}) is related to the signatures of the quantum state degradation. This is analyzed in Fig.  \ref{decoherence}, where we consider an adiabatic evolution of the QD from negative to positive energies, returning back to its original value at long times. Given $\left| 1,-1\right\rangle_i$ as the initial state for the MBSs, the quantum state of the system evolves after the first manipulation to a state $\propto V_Le^{-i\phi_L/2}\left| -1\,-1\right\rangle+V_Re^{i\phi_R/2}\left| 1,1\right\rangle$ for an adiabatic manipulation. To simulate the effect from decoherence, we include a random phase factor in one of the terms. The state before the second manipulation is thus given by $\propto V_Le^{-i\phi_L/2}\left| -1,-1\right\rangle+ V_Re^{i(\phi_R/2+\xi)}\left| 1,1\right\rangle$. The phase factor $\xi$ is taken as a random variable with all the values equally distributed within an interval. The result for different ranges of $\xi$ is shown in Fig. \ref{decoherence} (in the lower panel of Fig. 4 of the main text we have taken $\xi\in[-\pi/2,\pi/2]$), demonstrating the convergence to $\left\langle p_\nu\right\rangle=0$ for a case where the system coherence is completely lost in the process.

\end{document}